\documentclass[%
showpacs, 
 amsmath,amssymb,
 aps,
a4paper
]{revtex4}

\usepackage{graphicx}
\usepackage{dcolumn}
\usepackage{bm}
\usepackage{color}

\begin{document}

\preprint{}

\title{LHC updated hadronic interaction packages analyzed up to cosmic-ray energies}

\author{L. Calcagni}
\author{C. A. Garc{\'\i}a Canal}
\author{S. J. Sciutto}
\author{T. Tarutina}
\affiliation{%
	Departamento de F{\'\i}sica, Universidad Nacional de La Plata, and
	IFLP CONICET,
	C. C. 67, 1900 La Plata, Argentina}
%

\begin{abstract}
The results of high energy simulated experiments where a given hadronic particle impacts on a given target are statistically analyzed. The energy range of the projectiles goes from below the LHC scale up to the highest cosmic ray energies. This study was carried out by using the pre- and post-LHC versions of the hadronic interaction models QGSJET, EPOS and SIBYLL. Our analysis indicates that the post-LHC models present smaller differences in various quantities that characterize the secondary particles produced after the hadronic collisions, in comparison with the corresponding differences that are found comparing the respective old (pre-LHC) versions of the hadronic models. However, it is also found that there exist some discrepancies among models that persist even at the LHC energy scale, that call for further theoretical investigation.
An additional analysis of the impact that different modeling of hadronic collisions has on air shower development is also included. It consists of a detailed study of  the impact of the different pre- and post-LHC versions of the hadronic models considered, for relevant observables like the muon production depth distribution.
\end{abstract}

\maketitle

\section{Introduction}
\label{sec:level1}

One of the most challenging issues in contemporary research in physics of astroparticles is the determination of the chemical composition of the highest energy primary cosmic rays (CR). The composition estimations based on data generated during experimental observations require involved analysis techniques where a proper modeling of hadronic interactions constitutes an essential part of them. When passing through the Earth's atmosphere, the highly energetic CR interact with nuclei of air molecules and generate cascades of secondary particles, the extensive air showers (EAS). These secondary particles generated in EAS can then be detected and measured in some way and conclusions on the primary mass and energy can be drawn only after comparing them with the results of the EAS simulations. For this reason a proper modeling of the interactions that take place during the EAS development is essential for an adequate analysis.

Meanwhile the electroweak interactions are well understood, hadronic interactions, and especially their soft part, present substantial complications in their description. The observables of soft hadronic interactions are calculated using a combination of fundamental theoretical ideas based upon quantum chromodynamics (QCD) and empirical parametrizations. The differences between various implementations of hadronic interactions in the different available models constitute an important source of uncertainties in the determination of CR observables.

The effect on EAS observables of different modeling of hadronic interactions is the last consequence of a series of discrepancies between models. Previously, one can analyze the results of simulating single hadronic interactions and observe that differences are also present at this level, as has been reported in previous works \cite{Gar09,Lun04}.

A specific type of uncertainty comes from the needed model parameterizations that use particle accelerator data obtained for different (compared to CR) kinematic regions, energy range, and projectile-target configuration. The recently available data on proton-proton and proton-nucleus collisions at the Large Hadron Collider (LHC) have improved the knowledge of physics in an extended energy region with important consequences for the EAS simulations.  This new information was included in the new versions of many hadronic interaction packages, particularly in QGSJET-II  \cite{qgs24_1}, EPOS \cite{epolhc_1}, and SIBYLL \cite{sib23_1}.

The observables, important for EAS development, such as  multiplicity of secondary particles, inelasticity, fractions of secondary mesons and baryons, energy distributions of secondary particles and pseudorapidity distributions were analyzed in the already mentioned previous works \cite{Gar09,Lun04}, where the comparative analysis of the main hadronic packages available at the time was carried out. It was found that the different models present significant differences for all energy ranges. Moreover, in Ref. \cite{Gar09} special attention was given to study the so-called  very-energetic-leading-particle (VELP) events. These VELP events are characterized by (a) the small number of secondaries and (b) their leading particle carrying a substantial fraction of the projectile energy. Such events are very important in the shower development since they play a special role in transporting a significant fraction of the primary energy deep into the atmosphere. As a consequence, they are connected with the position of the EAS maximum $X_{\rm max}$, used to deduce the mass of primary CR particle. It was shown \cite{Gar09} that different models presented different results also for those VELP events.

The release of the mentioned updated versions of hadronic interaction models based on LHC recent data, motivated us to renew our study of the hadronic interaction packages. We find interesting to study the influence of the new experimental data included in the hadronic interaction models, on the EAS observables and on the characteristics of VELP events. 

We aim to discuss in this present work the coincidences or differences between the post-LHC hadronic interaction models and their impact on EAS observables and also the technical enhancements of each of the three packages since their previous versions. Whenever relevant, we compare our present  results with those obtained in our previous studies \cite{Gar09,Lun04}.

It is important to notice that a detailed analysis of the impact of diffractive interactions on EAS observables has been presented recently \cite{Arbe17}. In this work, the selection of diffractive events is done by means of changes in the settings of the corresponding internal parameters of the three different hadronic models considered, that permit enabling or disabling the diffractive interactions, instead of analyzing the properties of the produced secondaries to label the event as diffractive or not diffractive. The analysis of Ref. \cite{Arbe17} strongly support our findings related to the model dependence of the influence of diffractivelike interactions. 

In addition to the foregoing, there are also another analyses which compare the new post-LHC hadronic interaction models and their influence on the EAS observables \cite{Pierog:2017, ostap16} or which compare the hadronic packages with their previous version \cite{ostap13, epolhc_1, engel17}. Nonetheless, most of such studies focus almost entirely to illustrate how the models match the available LHC results and/or what is the impact on basic shower observables. Besides that, the present analysis includes a very detailed study of secondary production, fraction of VELP events and other related quantities that complete the mentioned works.

This work is organized as follows: in the next Sec.
\ref{sec:ModelTheory} we outline the main features of the hadronic
interaction packages QGSJET-II, EPOS and SIBYLL used in EAS
simulations comparing their previous versions with the updated
ones. In Secs. \ref{sec:Results} and \ref{sec:ShowerResults} we
discuss the obtained results of our hadronic model comparison, and the
impact on common shower observables, respectively. Our final remarks
and conclusions are placed in Sec. \ref{sec:Final}.

\section{Overview of the hadronic interaction models}
\label{sec:ModelTheory}

Perturbative quantum chromodynamics (pQCD) gives accurate results of hadronic production in high energy reactions when the processes are characterized by a large momentum transfer (large $Q^2$), the hard processes, such that the strong coupling $\alpha_{s}(Q^2)$ becomes small due to the asymptotic freedom property of QCD. However, high energy collisions involve mainly processes that are characterized by a small momentum transfer (soft processes), which escape the pQCD treatment.

To consider the non perturbative effects in QCD, and in order to describe soft hadronic interaction at high energies, the Gribov-Regge field theory (GRT) \cite{grib68} has been developed. In this approach, hadronic collisions are described as multiple scattering processes where in each of them, there is an exchange of a microscopic parton cascade. As in general one cannot use the pQCD description to treat such cascades, as most of these partons are characterized by small transverse momenta, they are treated phenomenologically as an exchange of an effective object, the Pomeron. The amplitude for the Pomeron exchange cannot be obtained from first principles and, therefore, it is introduced via a parametrization.

On the other hand, as the energy increases, a sizeable contribution of the so-called semihard processes appear. Such processes are characterized by a larger momentum transfer (compared to soft processes) so that pQCD can be applied, and results in the production of hadron jets with higher transverse momenta which can be observed.

In the GRT case,  as parton cascades contain both perturbative and nonperturbative parts, one can consider a general Pomeron as the sum of a soft and a semihard Pomeron, where the latter is represented as a QCD ladder between two soft Pomerons. This is the model adopted by the event generator packages  QGSJET-II and EPOS.

As an alternative to the Gribov-Regge theory, the QCD eikonal minijet approach has been used by the package SIBYLL. In this case, the hard sector is described by QCD ladder contributions  of the semihard Pomerons, while it is assumed that the soft interactions do not contribute significantly to the secondary particle production. The eikonal is based on the presence of minijets, where the particular features of high energy partonic interactions are described using the production of jets with low transverse momentum.  In earlier versions of SIBYLL, only one soft interaction was permitted. The last versions, SYBILL 2.3 and 2.3c, allow a larger range of phase space for soft interactions and adopts some aspects of GRT in order to accommodate multiple soft interactions, described as Pomeron exchanges.  

For particle production, SYBILL uses the Lund model of string hadronization where the transition from partonic entities to the final state hadron particles 
is accomplished by a massless relativistic string representing the QCD color force field. In this model the quarks and antiquarks are taken as string endpoint excitations and gluons as internal excitations of the color field. The phenomenon of quantum tunneling is responsible for the breakup of these strings and the appearance of new quark-antiquark pairs resulting in the hadronic remnant with excitation energy and momentum described by a phenomenological function. The new versions of SIBYLL consider also the possibility of gluon exchanges between sea quarks and sea and valence quarks and also allow for the break-up of diquarks. In its newest version, SIBYLL 2.3c \cite{sib2.3c}, the parameters of the fragmentation function, remnant excitation masses and string tensions were adjusted in order to obtain the correct Feynman scaling behavior.

In EPOS, two mechanisms of particle production are implemented, one is the decay of the hadronic remnant mentioned above and another is the hadronization of a cut Pomeron. A phenomenological Pomeron can be asociated with a QCD parton ladder attached to the projectile and target remnants with  multiparticle production resulting from the 
fact that the cut Pomeron can be seen as color strings with quarks (antiquarks) or diquarks ($\overline{qq}$) as string ends.  

QGSJET considers Pomeron-Pomeron interactions to take into account nonlinear effects related to parton shadowing and saturation. 
These interactions give rise to complex fan diagrams that are the source of particle production in this model. 

The results presented by the LHC correspond to collisions with center of mass energy around $\sqrt{s} \approx 10$ TeV. Consequently, for application in the highest energy cosmic rays physics it is necessary to extrapolate quantities like total proton-proton cross sections, for instance, at least one order of magnitude in energy (in the center of mass system). This extrapolation of data to higher energies is strongly dependent of the model used. 

In addition to that, it is necessary to calculate $\sigma_{p-air}^{tot}$ from $\sigma_{p-p}^{tot}$. To achieve this goal, the Glauber model \cite{glau59} is used.

As was already mentioned in the Introduction, one of the main sources of uncertainties in the numerical simulations with hadronic interaction packages is the unavailability of experimental data corresponding to the energy and kinematic region corresponding to EAS. Therefore the experimental data that recently became available from LHC is of prime importance for EAS physics. The energy reached in these LHC experiments (around 10 TeV in CM energy) correspond to the energies above the knee in the cosmic ray spectra, but still are about two orders of magnitude lower than that measured for example by the Pierre Auger Observatory \cite{AugerSpectrumIcrc17}.

The newest versions of the hadronic packages were tuned to reproduce the results of Run 1 of the LHC which are mainly the results of TOTEM \cite{totem1,totem2,totem3,totem4} and ATLAS \cite{atlas1} experiments. 
QGSJET-II-03 \cite{qgs23_0, qgs23_1, qgs23_2} updated to QGSJET-II-04 \cite{qgs24_1}, EPOS 1.99 \cite{epo99_1, epo99_2} changed into EPOS-LHC (v3400) \cite{epolhc_1} and SIBYLL 2.1\cite{sib21_1, sib21_2} upgraded to SIBYLL 2.3 \cite{sib23_1} and more recently to 2.3c \cite{sib2.3c}.

These new versions include adjustments considering the results of the measurement of total, elastic and inelastic proton-proton cross sections with high precision under various experimental conditions. This retuning of the models was able to eliminate many discrepancies between their predictions \cite{ostap16}. Also there are results for particle production in p-Pb by Alice \cite{alice1} and Pb-Pb collisions by ATLAS\cite{atlas2} and ALICE \cite{alice2} collaborations discussed in Ref. \cite{epolhc_1}, in connection to the new version of EPOS model, the EPOS-LHC.
  
It is important to mention that the newest version of SIBYLL, SIBYLL 2.3c, was adjusted to provide a better description of NA49 data. 
These data include  production of charged pions in p+p \cite{na491} and p+C interactions \cite{na492}; the production of protons, antiprotons and neutrons \cite{na493} and charged kaons \cite{na494} in p+p interactions; and the production of protons, antiprotons, neutrons, deuterons and tritons in minimum bias p+C interactions \cite{na495}.

The latest experiments in LHC make use of a large variety of forward detectors (see, for example, Ref. \cite{Berti:2017cfi} and references therein) to study events that are important for EAS development, those that we call VELP events (see Sec. \ref{sec:Results}), but to the best of our knowledge none of these data has still been taken into account to improve the hadronic interaction packages.

\section{Model Comparison Results}
\label{sec:Results}

We present here the results of the study of simulated experiments where a beam of given hadronic particles, the projectiles, impact on a given target undergoing hadronic collisions and generating secondary particles that are statistically analyzed. The input parameters of this simulated experiment are: (1) the type of primary particle, that can be a nucleon or a charged pion (other primaries could be also included but we restrict our present analysis to the mentioned ones); (2) the energy of the primary particle $E_P$; (3) the type of target, determined by its mass number $A$. For each parameter set, the collisions are simulated a sufficient number of times $N_{\rm ncoll}$, which in the present work is 10,000 unless otherwise specified.

We have run simulations for all the following combinations of: (1) hadronic models: QGSJET-II-03, QGSJET-II-04, EPOS 1.99, EPOS 3.4 (also known as EPOS-LHC),  SIBYLL 2.1, SIBYLL 2.3, SIBYLL 2.3c; (2) primary particles: protons, positive pions; and (3) targets: protons ($A=1$), and nitrogen nuclei ($A=14$). With this selection, we intend to cover the most relevant cases for the hadronic model comparison. In the case of the  targets, we want to mention that the selection of nitrogen is due to the fact that this nucleus is the most abundant in the Earth's atmosphere and it is therefore a representative case for the simulation of hadronic collisions that take place in such medium. On the other hand, the use of proton targets allows for studying the characteristics of collisions that take place in similar conditions as the real experiments whose data has been used to tune the models, for example proton-proton collisions at LHC. The energies of the projectiles range from 100 GeV (a typical energy near the threshold energy for all the considered models) up to about 300 EeV (corresponding to the highest cosmic ray energies observed).

\begin{figure}[p]
	\includegraphics{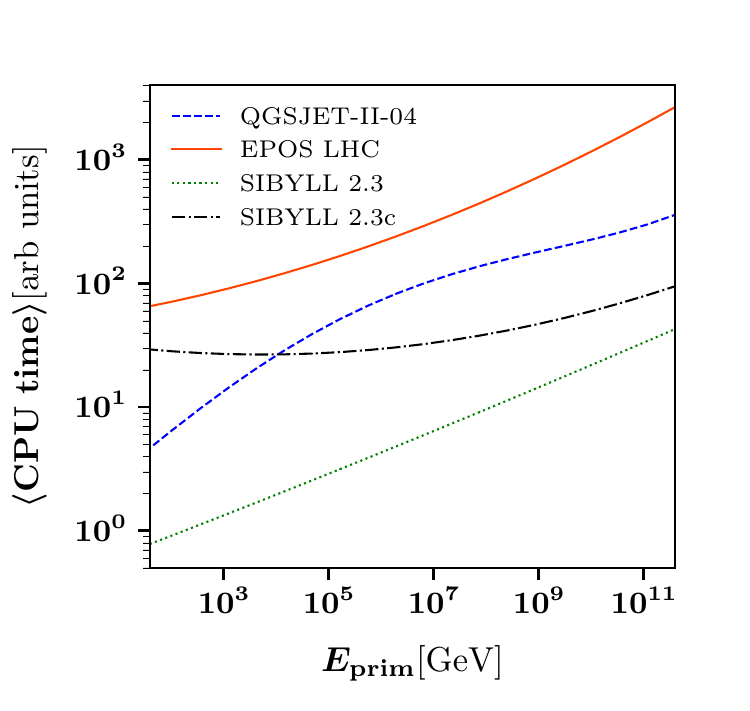}
	\caption{Average processor time required per collision versus primary energy, in the case of proton-nitrogen collisions.}
	\label{fig:CPUTvsEprim}
\end{figure}

To start with the analysis of our simulations results, let us address the technical question of the processing time required by each one of the hadronic models used. The most outstanding characteristic in this sense is the enormous difference of processing time requirements of each package. In Fig. \ref{fig:CPUTvsEprim} the average processing time is plotted as a function of the primary energy, in the case of proton-nitrogen collisions, and considering the newest versions of the QGSJET, EPOS and SIBYLL models. The vertical scale is normalized taking as 1 the average processing time required by SIBYLL 2.3 to process a 300 GeV p-N collision. It shows up clearly from this figure that SIBYLL 2.3 is the fastest collision generator, while EPOS is the slowest one, with processing time requirements that in some cases are more than two orders of magnitude larger in comparison with SIBYLL. QGSJET requirements are also large in comparison with SIBYLL 2.3, but remain in all cases smaller that those of EPOS.

Needless to say, such important differences are most probably due to the fact that the models here considered process the collisions using algorithms with different degrees of theoretical and computational complexity. Our main interest is to report on the processing time consumption from the point of view of the normal user of an air shower simulation program, thus skipping any detailed analysis about the characteristics of the models internal algorithms \footnote{The differences in CPU time for the different hadronic models could be related with the fact that SIBYLL, at difference with QGSJET and EPOS, does not include DGLAP evolution for hard scattering and that EPOS includes energy sharing at amplitude level and collective effect, of clear interest for heavy ion collisions. We acknowledge T. Pierog for clarifying this point to us.}.

The old versions of both EPOS and QGSJET require processing times that are very similar to the corresponding one for the newest versions plotted in Fig. \ref{fig:CPUTvsEprim}, and have therefore not been plotted for clarity. On the other hand, in the case of SIBYLL we included in Fig. \ref{fig:CPUTvsEprim} the last two versions of this hadronic package since there is a noticeable increase in processing time when comparing the recently released version 2.3c with the previous one 2.3. This difference is particularly large (more than one order of magnitude) at small primary energies.

It is also important to mention that the processing time required for a given collision spreads very widely around the mean values plotted in Fig. \ref{fig:CPUTvsEprim}. As an example of this characteristic of the processing time distribution, we observe that in the case of 100 EeV collisions, the processing time for EPOS (SIBYLL 2.3) can overpass in more than 60 (30) times the corresponding average.

\begin{figure}[p]
	\includegraphics{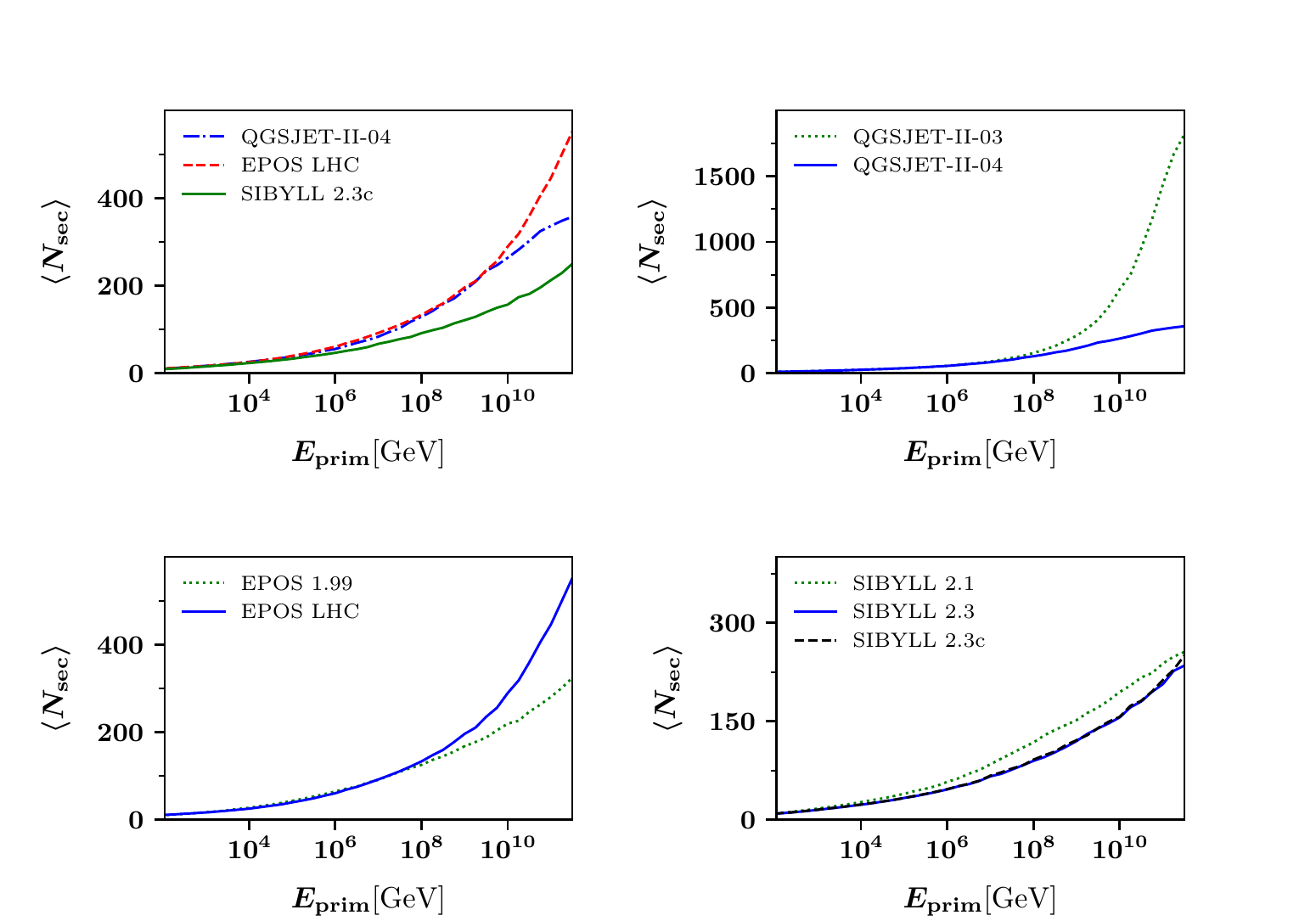}
	\caption{Average number of secondary particles, $\langle{N_{\rm sec}}\rangle$, versus primary energy for the case of proton-proton collisions.}
	\label{fig:NsecppvsEprim}
\end{figure}

\begin{figure}[p]
	\includegraphics{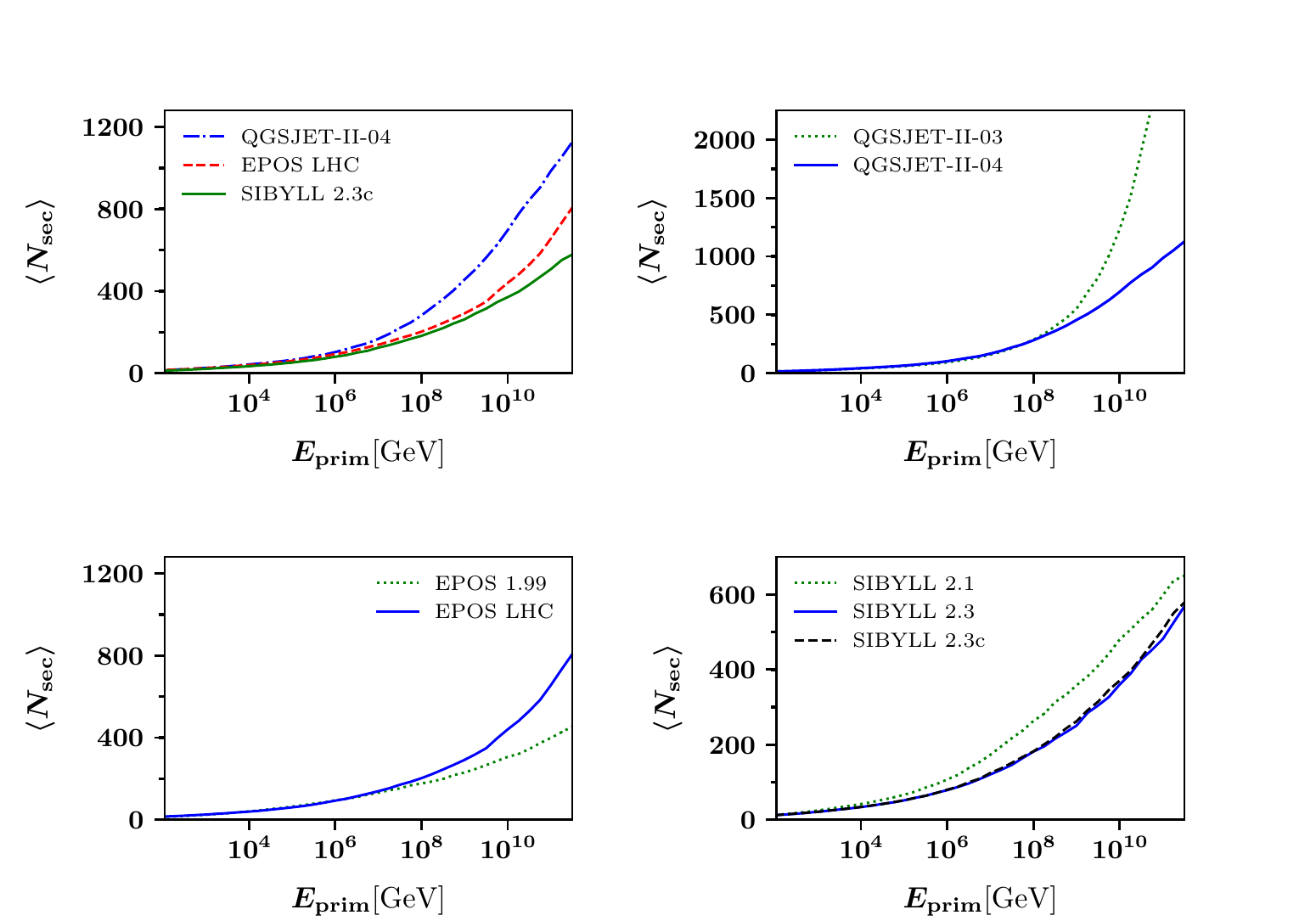}
	\caption{Same as figure \ref{fig:NsecppvsEprim}, but for the case of proton-nitrogen collisions.}
	\label{fig:NsecpNvsEprim}
\end{figure}

The number of secondaries, $N_{\rm sec}$, produced after an hadronic collision is normally the first observable to be analyzed.
In Figs. \ref{fig:NsecppvsEprim}-\ref{fig:NsecpNvsEprim} we present the dependence of $\langle{N_{\rm sec}}\rangle$ with the primary energy for the cases of proton-proton and proton-nitrogen collisions, respectively. Both figures include a comparison among the new versions of the different hadronic models (upper left plots) and also each model with its corresponding previous version. In both figures it can be seen that for the new versions of the different models, SIBYLL 2.3 produces the smallest number $\langle{N_{\rm sec}}\rangle$. This is in agreement with the fact that one of the main  differences between the semi-hard Pomeron scheme and the minijet approach employed in SIBYLL is that in the former case there is an additional contribution to secondary particle production which emerges from the soft parton evolution.

In the case of proton-nitrogen collisions, the largest number of $\langle{N_{\rm sec}}\rangle$ corresponds to QGSJET-II-04, then followed by EPOS-LHC. This behavior is slightly different with respect to the case of proton-proton collisions where the largest number of $\langle{N_{\rm sec}}\rangle$ corresponds to EPOS-LHC, then followed by QGSJET-II-04. The smaller average number of secondaries predicted by EPOS-LHC with respect to QGSJET-II-04 is consistent with the results of \cite{Ostapchenko:2014mna}.

When considering the case of proton-proton collisions (Fig. \ref{fig:NsecppvsEprim}), it can be seen that the mean number of secondary particles are significantly lower than the corresponding ones for the proton-nitrogen case. For both targets (Figs. \ref{fig:NsecppvsEprim}-\ref{fig:NsecpNvsEprim}) there are lower values of $\langle{N_{\rm sec}}\rangle$ at high energies for the new models in comparison with their previous version, except for the case of EPOS.

Notice also that the ratios between the average numbers of secondaries displayed in the upper left plots of Figs. \ref{fig:NsecppvsEprim}-\ref{fig:NsecpNvsEprim} are not the same for proton-proton or proton-nitrogen collisions. Furthermore, in the case of proton-proton collisions the average secondaries of QGSJET and EPOS are virtually coincident for energies up to nearly $10^{10}$ GeV, while in the case of proton-nitrogen collisions the coincidence is now between SIBYLL and EPOS curves.

Following our previous work \cite{Gar09}, we classify all collision events as being either ``VELP'' (very energetic leading particle) events, or simply ``inelastic'' events. The interest for such a classification is closely related with the study of hadronic collisions in the framework of particle showers that develop in the Earth's atmosphere. The algorithm for labeling an event as ``VELP''  or ``inelastic'' used in the present work, described in detail in Ref. \cite{Gar09}, allows one to determine whether or not a collision event contains an energetic leading particle capable of contributing considerably to the energy transport deep down in the atmosphere during the air shower development. One of the quantities that are considered in the mentioned algorithm is the {\em leading energy fraction\/} \cite{Gar09}, $f_l$, defined as
\begin{equation} \label{eq:flkdef}
f_l = 1 - K =\frac{E_{\rm lead}}{E_P}
\end{equation}
where $E_{\rm lead}$ is the energy of the most energetic secondary emerging from the collision (leading particle). The complementary quantity $K$ is the {\em inelasticity.\/}
In the case of a VELP event, $f_l$ is close or very close to 1, or, equivalently, $K$ close or very close to 0. Non VELP, i.e., inelastic events, will present wide distributions of $f_l$ or $K$.

It is worth mentioning that VELP events certainly include most of the standard diffractive events \cite{Gar09}. Consequently, the fraction of VELP events, defined as the ratio between the number of VELP events divided by the total number of events, gives an estimation of the diffractive to total cross section ratio.

\begin{figure}[p]
	\includegraphics{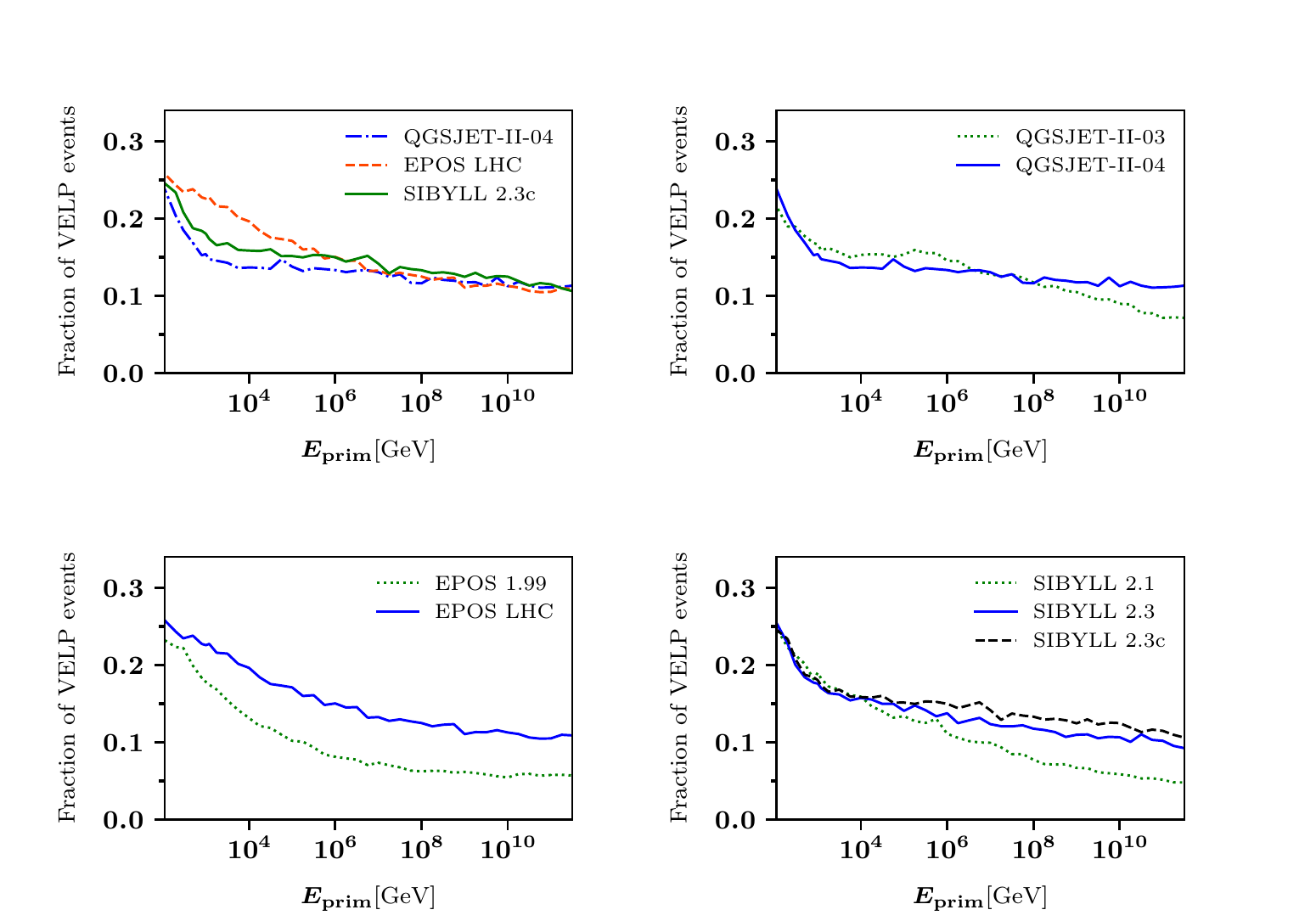}
	\caption{Fraction of VELP events versus primary energy for the case of proton-proton collisions.}
	\label{fig:FracVelppp}
\end{figure}

\begin{figure}[p]
	\includegraphics{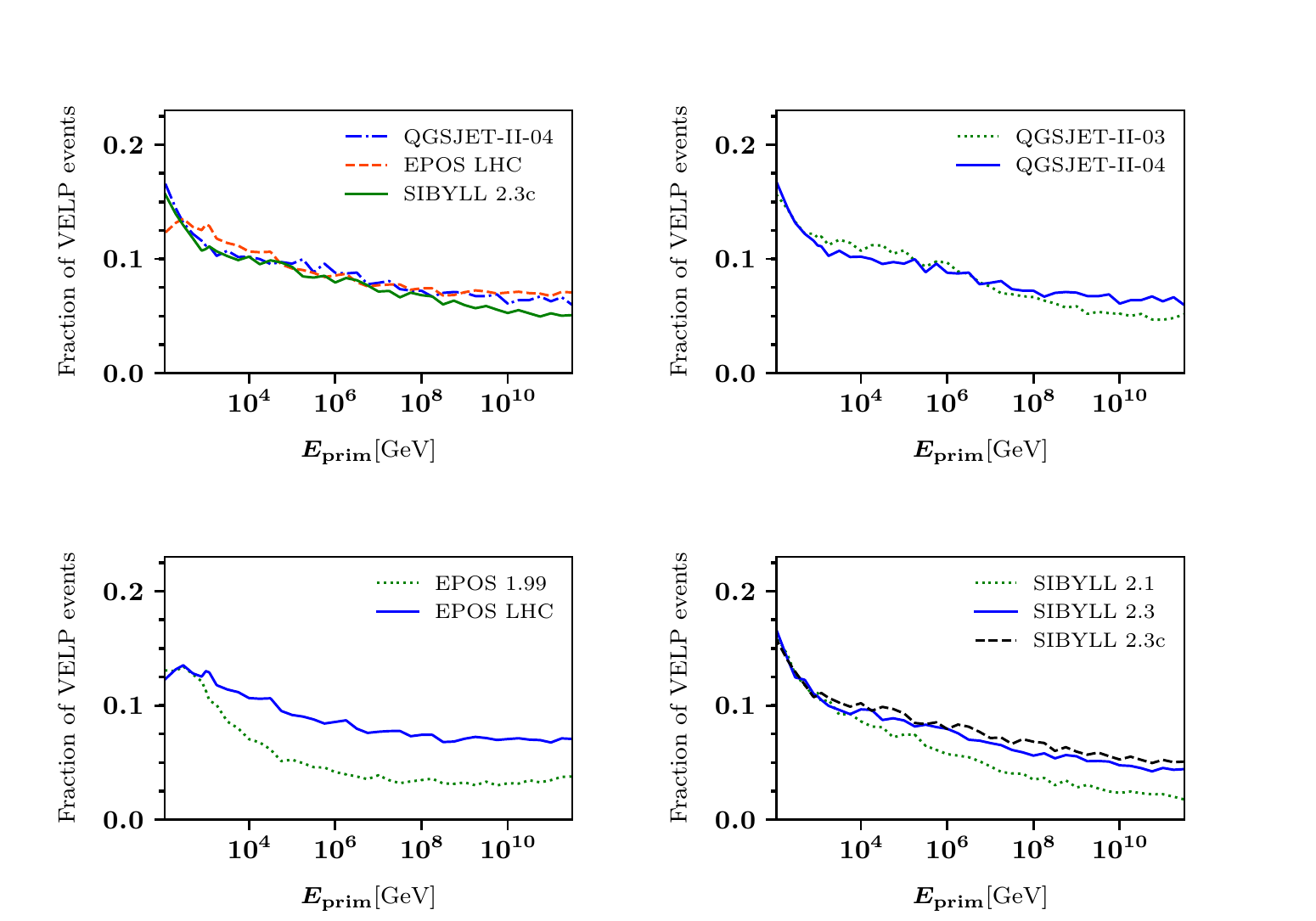}
	\caption{Same as figure \ref{fig:FracVelppp}, but for the case of proton-nitrogen collisions.}
	\label{fig:FracVelppN}
\end{figure}

In Figs. \ref{fig:FracVelppp} and \ref{fig:FracVelppN} we present the dependence of the fraction of VELP events with the primary energy for the cases of proton-proton and proton-nitrogen collisions, respectively. In the same way as in the cases of Figs. \ref{fig:NsecppvsEprim}-\ref{fig:NsecpNvsEprim}, we compare the new models (upper left plots), and also each model with its corresponding previous version.

As VELP events are characterized by a low number of secondary particles, the plots in Figs. \ref{fig:FracVelppp} and \ref{fig:FracVelppN} are inversely related to the respective ones in Figs. \ref{fig:NsecppvsEprim} and \ref{fig:NsecpNvsEprim}. There is a larger number of VELP events for the cases of proton-proton collision in comparison with the proton-nitrogen collisions.

For most of the analyzed energy range, post-LHC models give rise to a larger fraction of VELP events, particularly at the highest energies, than the ones corresponding to the respective pre-LHC versions. The increment in the fraction of VELP events is clearly noticeable in the cases of SIBYLL and EPOS and for both of the studied targets (protons and nitrogen nuclei), as can be seen from Figs. \ref{fig:FracVelppp} and \ref{fig:FracVelppN}. In the case of QGSJET such increment is not so important.   When comparing the newest versions of the hadronic models (upper left plots of Figs. \ref{fig:FracVelppp} and \ref{fig:FracVelppN}), it can be seen that in the case of proton-nitrogen collisions the models return similar figures for the entire energy range under analysis, with the exception of SIBYLL that presents a particularly small fraction of VELP events at the highest energies. On the other hand, in the case of proton-proton collisions, all the models return similar results at the highest energies, while at energies below 1 PeV SIBYLL and QGSJET predictions also agree, in contrast with EPOS that in this primary energy range returns noticeably larger fractions of VELP events.

Notice also that in the case of proton-proton collisions, and at LHC energies (from 32 to 85 PeV in the lab reference system), the three models return similar fractions of VELP events.

\begin{figure}[p]
	\includegraphics{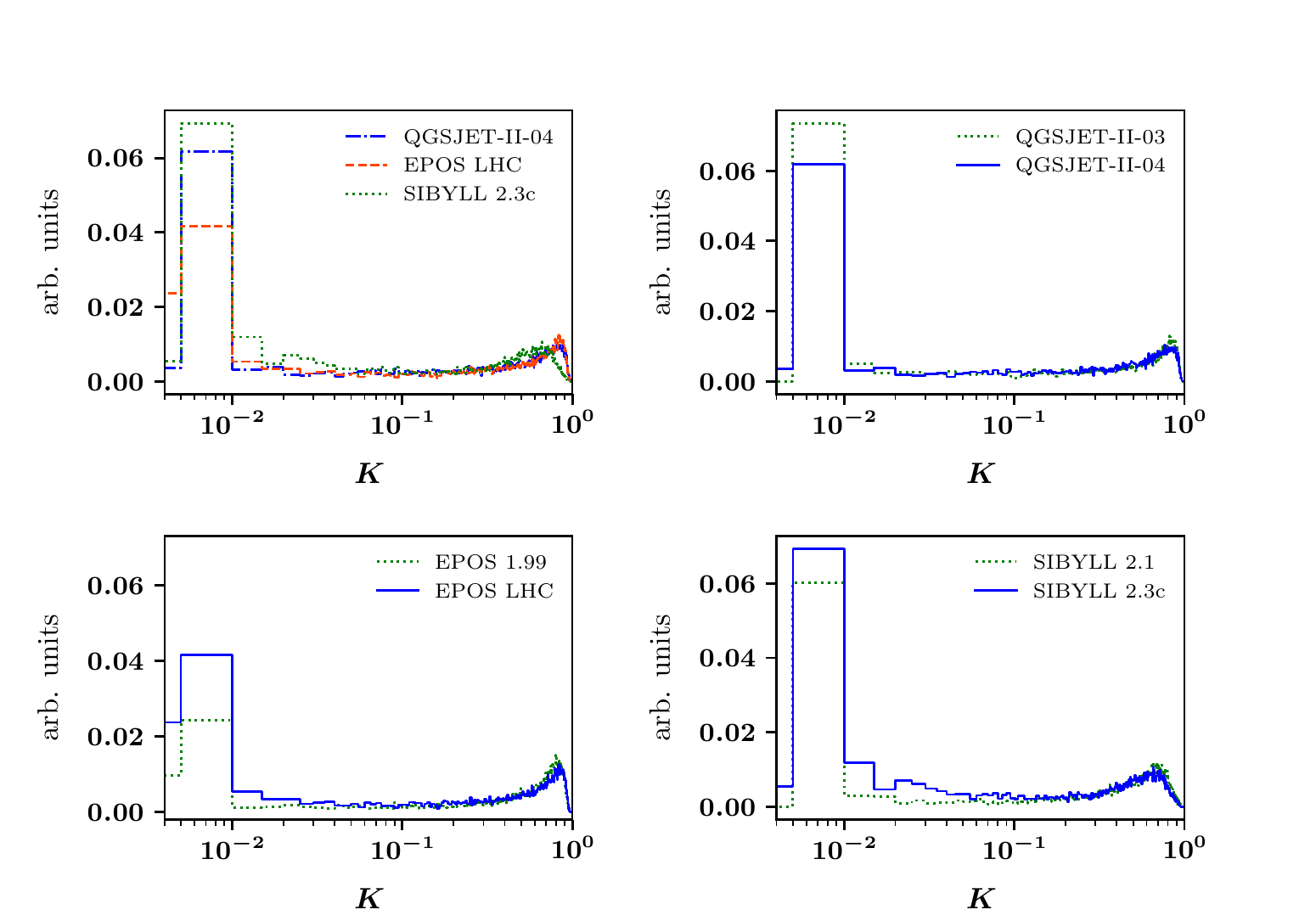}
	\caption{Distribution of the inelasticity in the case of  56 PeV proton-proton collisions.}
	\label{fig:Kdist56PeVpp}
\end{figure}

\begin{figure}[p]
	\includegraphics{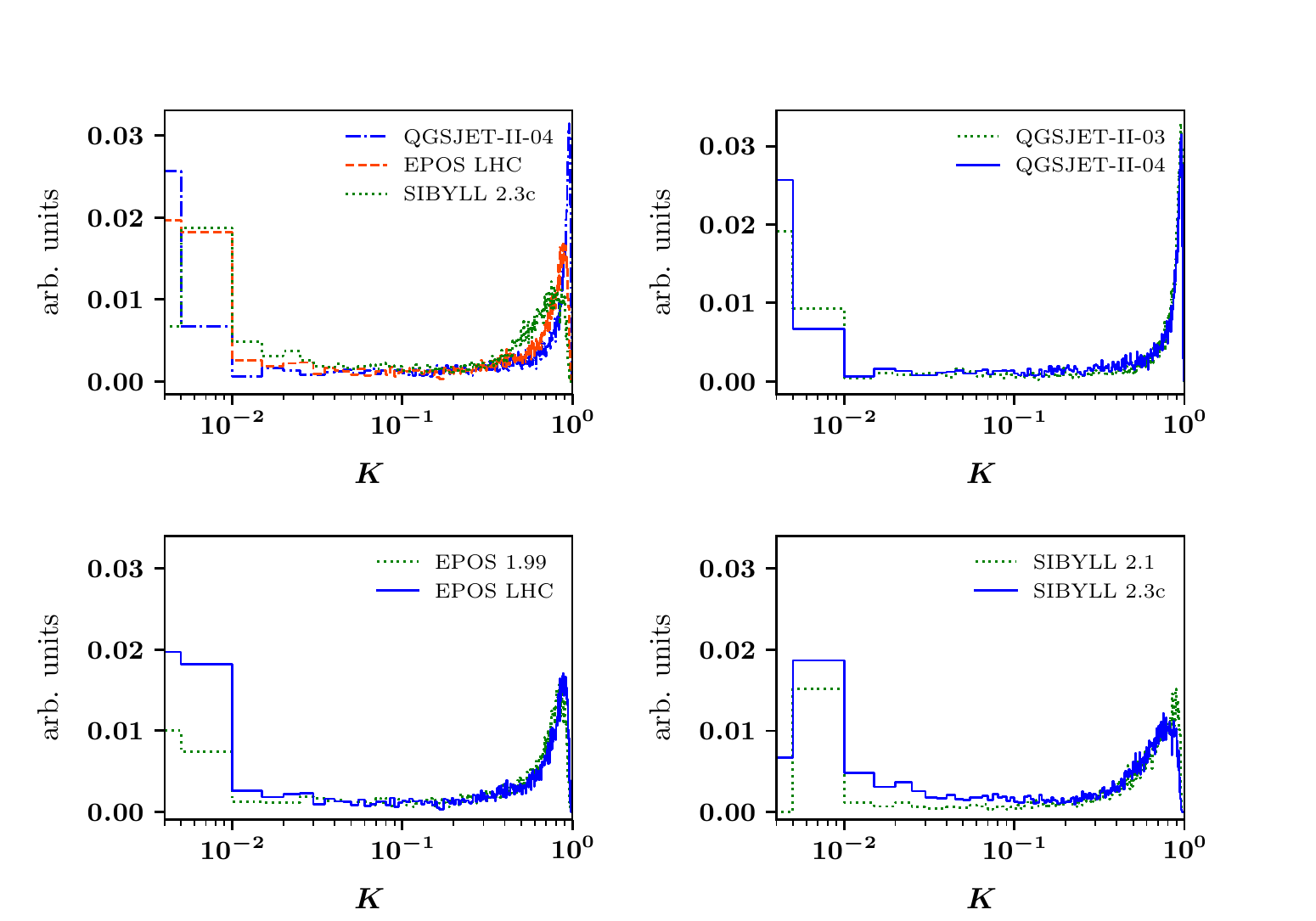}
	\caption{Same as figure \ref{fig:Kdist56PeVpp}, but for 100 EeV proton-nitrogen collisions.}
	\label{fig:Kdist100EeVpN}
\end{figure}

The analysis of other observables allows one to obtain a more complete picture of the similarities and differences between hadronic models, and for this reason we have also analyzed the inelasticity (Eq. \ref{eq:flkdef}). In Fig. \ref{fig:Kdist56PeVpp} (\ref{fig:Kdist100EeVpN}) the inelasticity distributions for 56 PeV proton-proton (100 EeV proton-nitrogen) collisions are presented for the cases of the three models studied.

The differences between the distributions corresponding to pre- and post-LHC versions of the models are in general not large, as can be seen in Fig. \ref{fig:Kdist56PeVpp}. In the case of events with a very small inelasticity ($f_l$ near 1), one finds that their frequency is larger with the newer versions of SIBYLL and EPOS. This is consistent with the increase in the fraction of VELP events that can clearly be seen in the plots of figure \ref{fig:FracVelppp}. The very similar QGSJET inelasticity distributions are also consistent with the data displayed in Fig. \ref{fig:FracVelppp}.

The comparison of the inelasticity distributions corresponding to the newer versions of the three analyzed models (upper left plot of Fig. \ref{fig:Kdist56PeVpp}), shows that all the distributions are very similar, but not completely coincident, in contrast with the fact that the corresponding fractions of VELP events are virtually the same (see upper left plot of Fig. \ref{fig:FracVelppp}).

\begin{figure}[p]
	\includegraphics{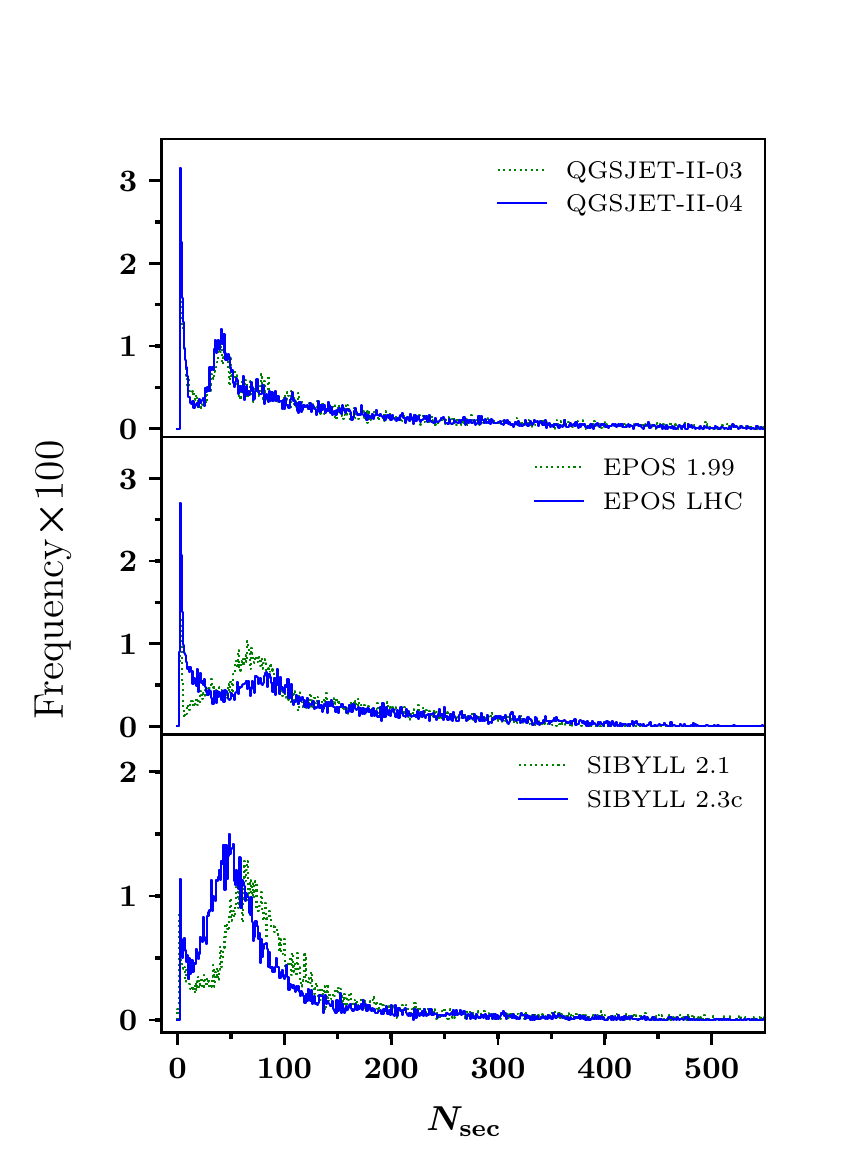}
	\caption{Distribution of the number of secondaries, $N_{\rm sec}$, for 56 PeV
		proton-proton collisions. The vertical scale of frequencies corresponds to fractions of the total number of simulated events.}
	\label{fig:Nsecpp56PeV}
\end{figure}

\begin{figure}[p]
	\includegraphics{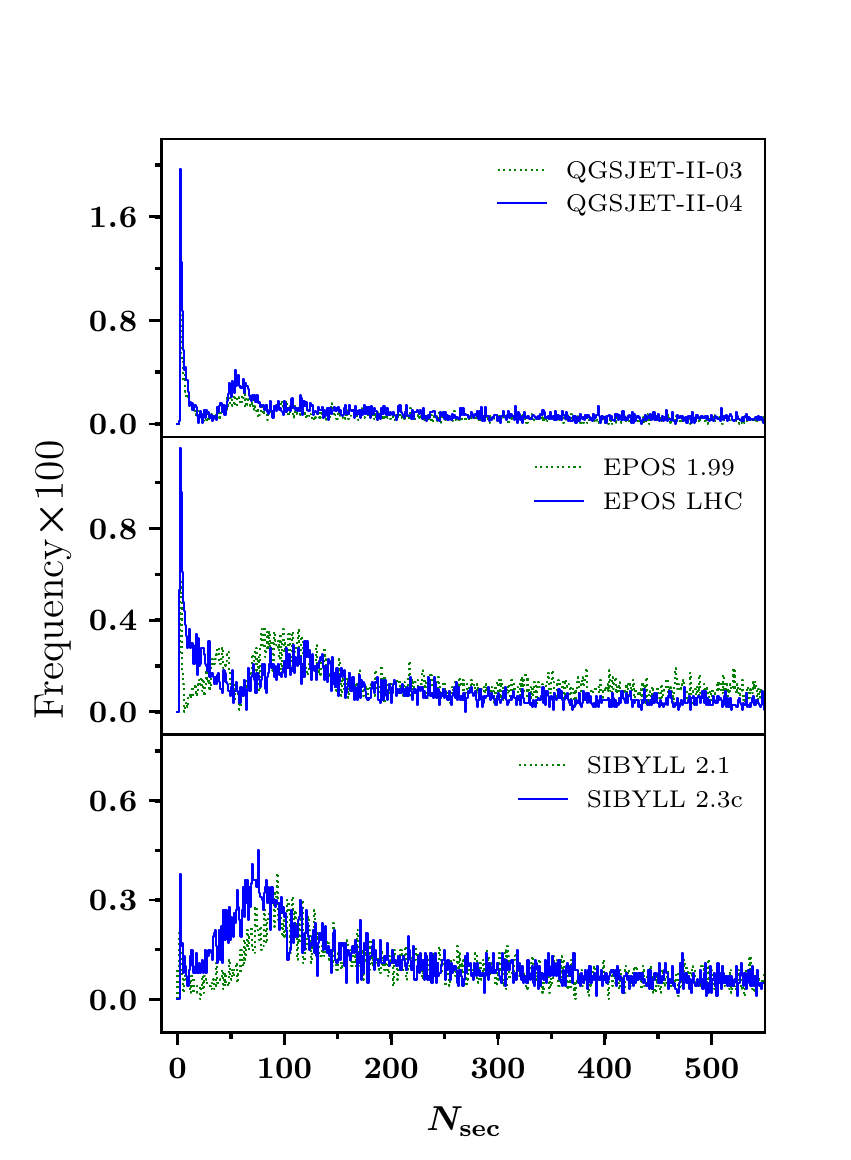}
	\caption{Same as figure \ref{fig:Nsecpp56PeV}, but for 100 EeV proton-nitrogen collisions.}
	\label{fig:NsecpN100EeV}
\end{figure}

To better understand this situation, we include in Figs. \ref{fig:Nsecpp56PeV} and \ref{fig:NsecpN100EeV} the distributions of $N_{\rm sec}$ for 56 PeV proton-proton and 100 EeV proton-nitrogen collisions, respectively, showing a comparison between pre- and post-LHC models. In all cases a frequency peak at low $N_{\rm sec}$ shows up clearly, which corresponds mainly to VELP events. The QGSJET distributions present virtually no differences when comparing its pre- and post-LHC versions. On the other hand, the EPOS distributions show important qualitative differences when comparing the pre -and post-LHC cases. In the newest version of EPOS a peak of the distribution for events with 4 or 5 secondary particles is clearly noticeable. Such peak is larger than the corresponding one for EPOS 1.99. This is consistent with the fact that EPOS-LHC has larger fractions of VELP events, as shown in Figs. \ref{fig:FracVelppp} and \ref{fig:FracVelppN} (lower left plots). However, the EPOS 1.99 distribution presents a larger peak around the 75 secondary particles zone. Such events have low inelasticity. As a result, the number of events with low inelasticity is not much different from the corresponding one for EPOS-LHC, as can be seen from Figs. \ref{fig:Kdist56PeVpp} and \ref{fig:Kdist100EeVpN}. In the case of SIBYLL, the qualitative structure of the distribution of the number of secondary particles is similar in the pre- and post-LHC cases, and presents an important peak in the region that spans events with about 30 to 150 secondary particles. Finally, QGSJET in both its pre- and post-LHC versions also presents such a peak, smaller in comparison with the sharp few-event peak.

Another very important characteristic to analyze is the kind of secondary hadrons that these models produce as output after each simulated collision. It is important to take into account here that the whole process of building the final list of secondaries encompasses the steps of energy splitting, hadron creation, and eventual decay of unstable hadrons or resonances. The hadronic models that we have studied allow the user to control what particles are considered ``unstable", with the exception of QGSJET where we have not found user-controllable parameters to control decay of unstable particles. When such a secondary particle is created it undergoes further processing being forced to decay. As a result, the output secondary particle list does not include the unstable particle, but rather its decay products. In our study, we have configured the hadronic packages similarly as they are used within common air shower simulation programs, using the default settings that in general force to decay only very short lived resonances. For this reason, our analysis of the kind of secondaries produced refers always to the final list of secondary particles effectively coming off after every collision is processed, without distinguishing between hadrons created by the collision processing engine or particles that are product of decays that were processed internally.

\begin{figure}[pt]
	\includegraphics{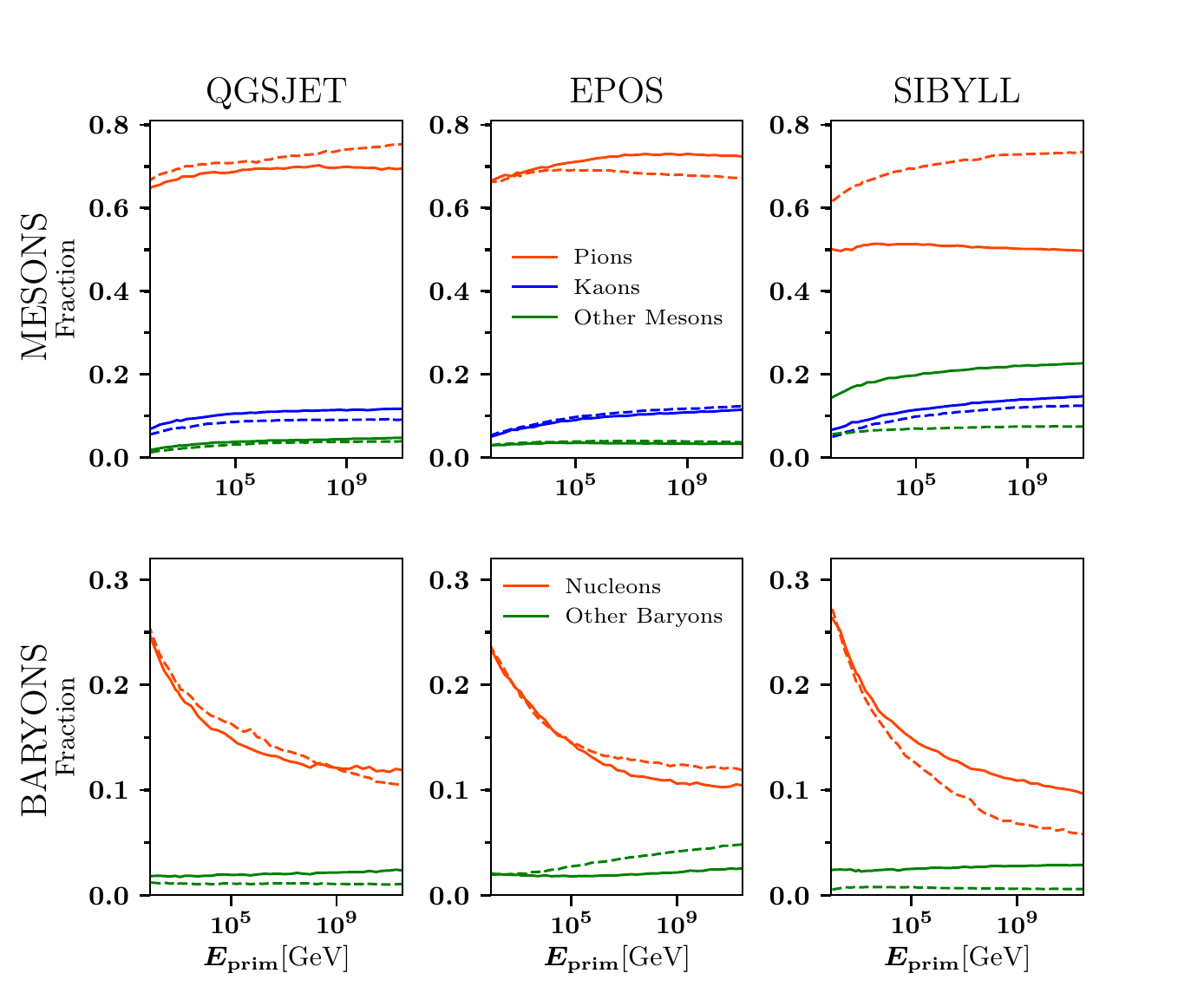}
	\caption{Fraction of secondary $\pi$, $K$, and other mesons (upper row), and nucleons and other baryons (lower row) versus
		primary energy for the case of proton-proton collisions, and for each hadronic model.  The plots in the left, middle, and right side columns correspond to the QGSJET, EPOS, and SIBYLL models, respectively.
		The post-LHC (old) models are represented with solid (dashed) lines.}
	\label{fig:SecFractsppvsEprim}
\end{figure}

\begin{figure}[tp]
	\includegraphics{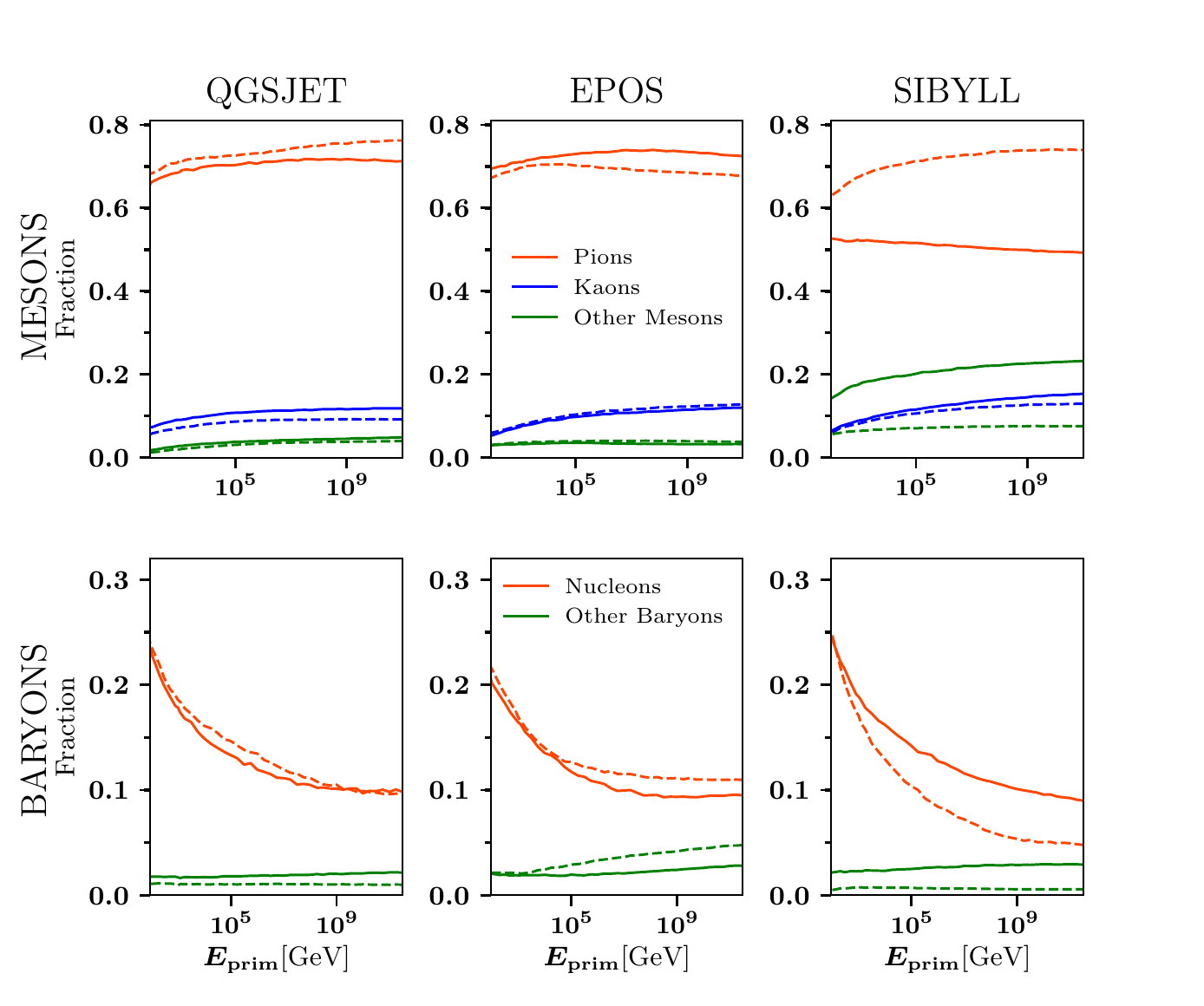}
	\caption{Same as figure \ref{fig:SecFractsppvsEprim}, but for the case of proton-nitrogen collisions.}
	\label{fig:SecFractspNvsEprim}
\end{figure}

To start with this analysis of the kind of secondary particles that emerge after each collision, let us refer to the plots of Figs. \ref{fig:SecFractsppvsEprim} and \ref{fig:SecFractspNvsEprim}. In such figures, the average fractions of the most relevant groups of hadrons produced after the collisions are plotted as functions of the primary energy. The groups of particles considered are: pions (charged and neutral), kaons (charged and neutral), other mesons (mainly $\eta$ and $\rho$), nucleons ($p$, $n$, $\bar p$, $\bar n$), and other baryons (mainly $\Lambda$). 

QGSJET-II-04 and EPOS LHC present a similar behavior for the fraction of each type of secondary particle in comparison with their previous version for both proton-proton and proton-nitrogen cases. However, in the case of SIBYLL, there is an appreciable decrement of the fraction of pions and an increment of the fraction of ``other mesons" in the new version of this hadronic model. Also, at high energies there is an increment of the fraction of nucleons and ``other baryons". These differences can be understood taking into account that the last version of SIBYLL (2.3c) extends the fragmentation model to increase baryon pair production and also includes the production of charmed hadrons \cite{Rie15}.

\begin{figure}[pt]
	\includegraphics{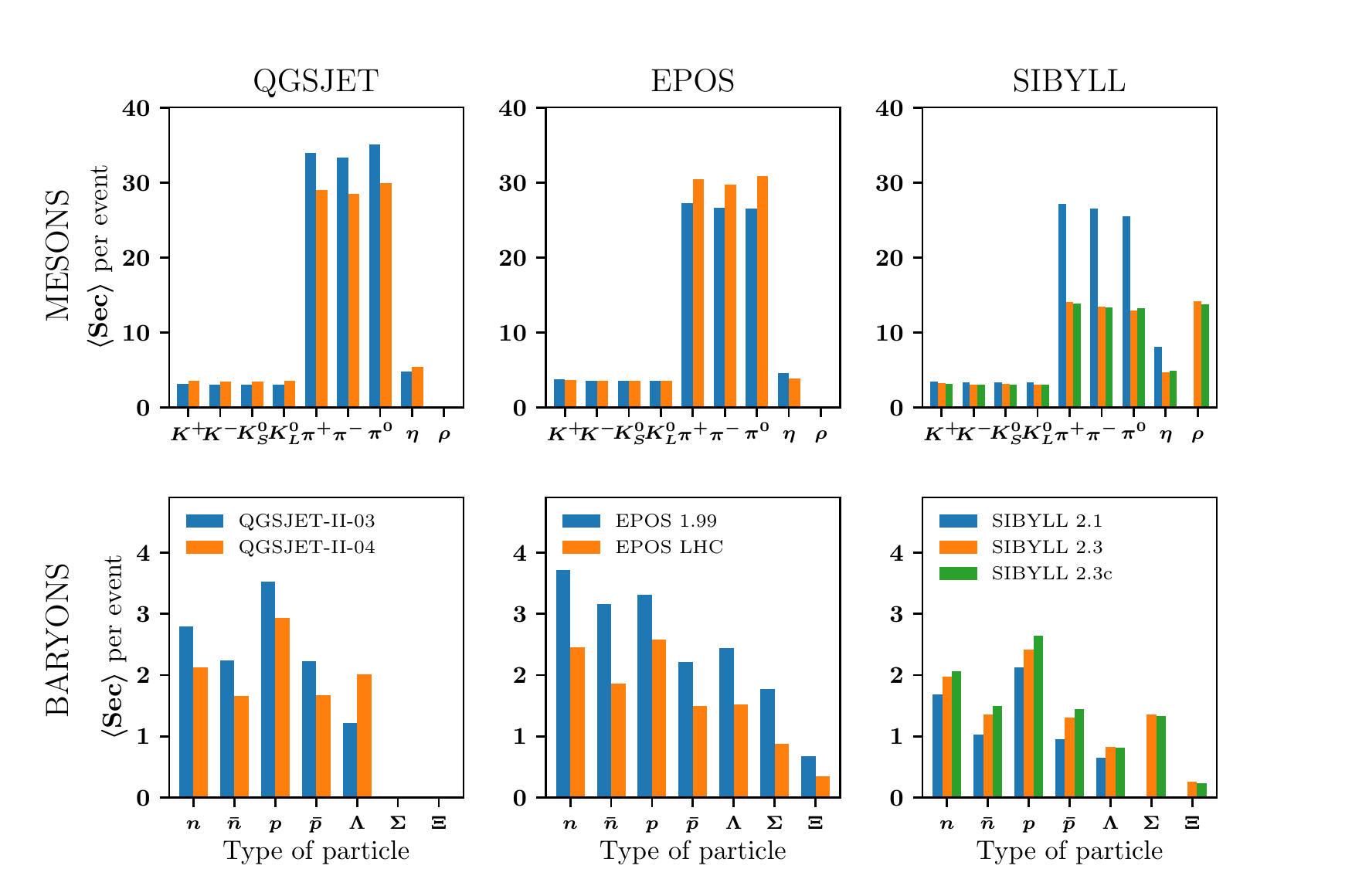}
	\caption{Distribution of secondary mesons and baryons for the case of 56 PeV proton-proton collisions.}
	\label{fig:SecDistpp56PeV}
\end{figure}

\begin{figure}[pt]
	\includegraphics{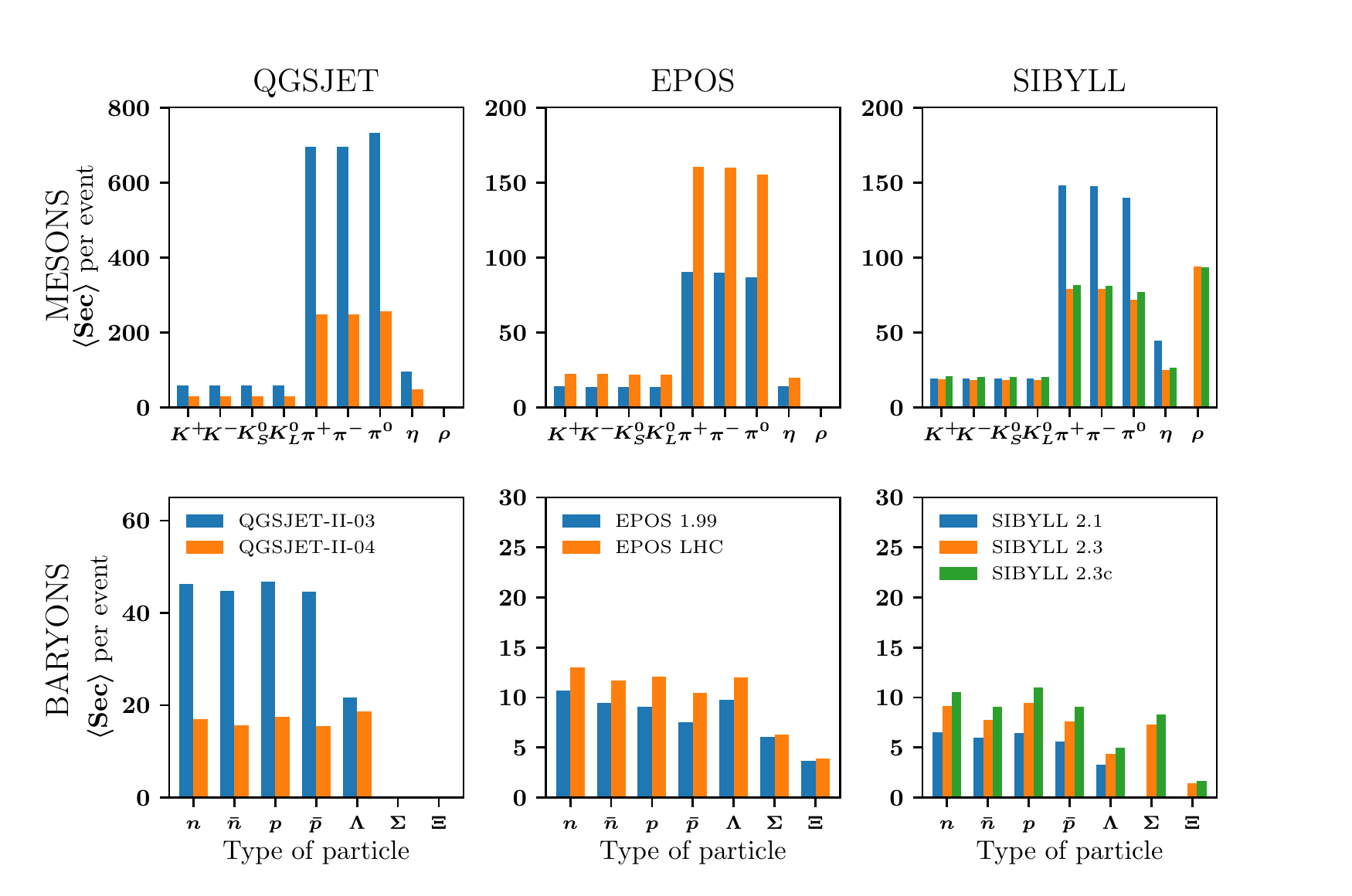}
	\caption{Distribution of secondary mesons and baryons for the case of 100 EeV proton-nitrogen collisions.}
	\label{fig:SecDistpN100EeV}
\end{figure}

All these changes in the fractions of secondary particles can be better appreciated in Figs. \ref{fig:SecDistpp56PeV} and \ref{fig:SecDistpN100EeV} that display the distributions of the average number of the most relevant secondary mesons and baryons for 56 PeV proton-proton and 100 EeV proton-nitrogen collisions respectively. We show the results for the pre- and post-LHC versions of the different hadronic interaction models.

To improve the visual aspects of the plots, in  Figs. \ref{fig:SecDistpp56PeV} and \ref{fig:SecDistpN100EeV} the particles included are the same for all the considered hadronic models, allowing for zero length bars in the cases where such particles are never present among the secondaries that emerge off the collisions: $\rho$'s, $\Sigma$'s, and $\Xi$'s for QGSJET; and $\rho$'s for EPOS. 

In both proton-proton and proton-nitrogen cases it can be seen that QGSJET produces the lowest variety of baryons. QGSJET returns the largest numbers of secondaries, particularly at the highest energies, requiring that a different vertical scale is needed for this model in these plots, for both mesons and baryons. On the other hand, we use the same scale for EPOS and SIBYLL in all cases.

QGSJET shows noticeably smaller numbers of mesons and baryons in comparison with its older pre-LHC versions, in agreement with the data displayed in Figs. \ref{fig:NsecppvsEprim}-\ref{fig:NsecpNvsEprim} (upper right plots). In the case of EPOS, there is a noticeable increase in the number of secondary mesons, particularly pions, when passing from the old (1.99) to the LHC version. However, both versions return similar numbers of baryons.

On the other hand, SIBYLL 2.1 returns a substantially larger number of pions in comparison with the recent 2.3 or 2.3c versions. This is compensated with the production of $\rho$'s resulting in a small variation in the total number of mesons when comparing pre- and post-LHC versions of SIBYLL. Notice that SIBYLL 2.1 does not produce $\rho$'s. In the case of baryons, the production of neutrons, protons and $\Lambda$'s and their antiparticles is larger in the recent versions of SIBYLL, particularly in SIBYLL 2.3c. We would like to mention that the recent versions of SIBYLL give a very good description of $\rho^0$ production at 350 GeV \cite{NA61SHINE17}.

It is important to mention that the configuration we used for these runs corresponds, as it has already been mentioned, to the default set of parameters employed in common air shower simulation programs. In the case of EPOS (all versions), such setting implies that $\rho$ mesons are internally forced to decay. For this reason there are no such particles in the EPOS plots of Figs. \ref{fig:SecDistpp56PeV} and \ref{fig:SecDistpN100EeV}. Should these forced decays be disabled, both versions of EPOS would output a significant number of $\rho$ mesons in the two cases here considered: for 56 PeV proton-proton collisions the EPOS $\rho$ production is slightly smaller that the here reported figures for SIBYLL 2.3, and reduces to approximately 50\% for 100 EeV $p$-N collisions.

Notice also that EPOS produces a very small number of $\Omega$'s, about 2 (30) every 100 events in the conditions of Fig. \ref{fig:SecDistpp56PeV} (\ref{fig:SecDistpN100EeV}). Such kind of particles are not present among the secondary particles generated with SIBYLL or QGSJET.

It is also worth mentioning that among all the secondaries that are produced by EPOS or SIBYLL (in all of their versions) there can be a small number of photons, leptons, and neutrinos, which come from decays that were processed internally by the corresponding simulation packages. We recall that there is no such kind of particles within the secondaries generated by QGSJET.

\begin{figure}[pt]
	\includegraphics{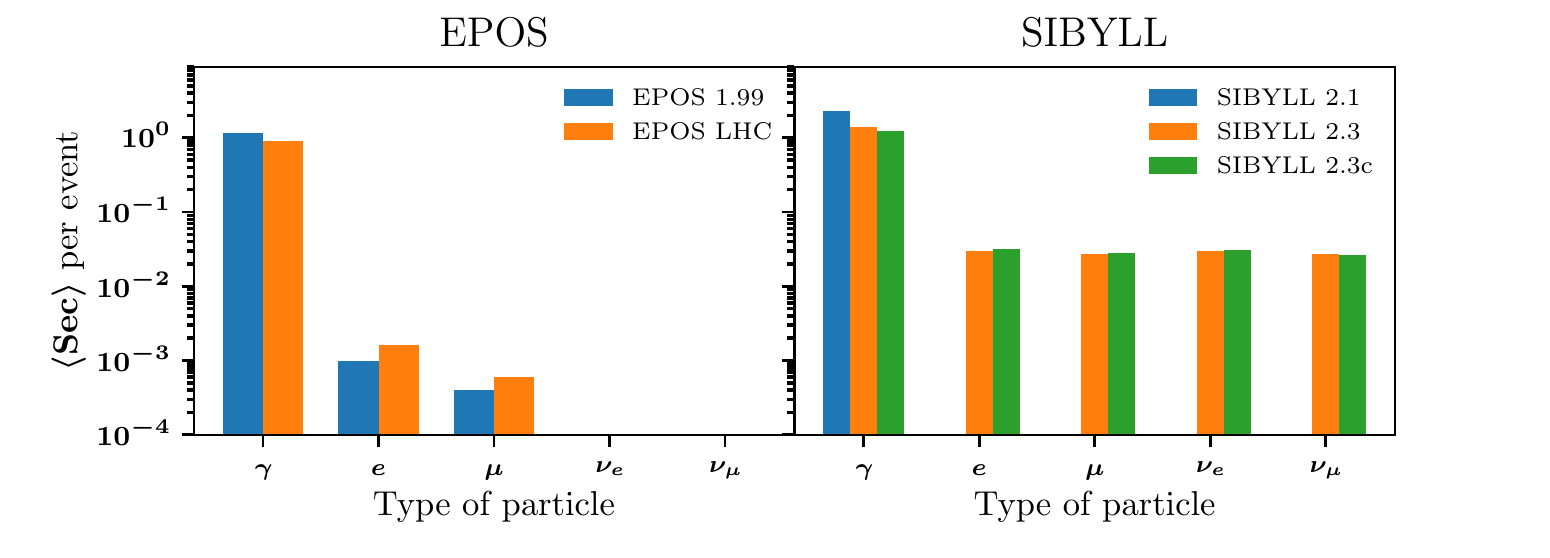}
	\caption{Distribution of secondary photons, leptons and neutrinos for the case of 56 PeV proton-proton collisions.}
	\label{fig:SecDistpp56PeV_lepgam}
\end{figure}

\begin{figure}[pt]
	\includegraphics{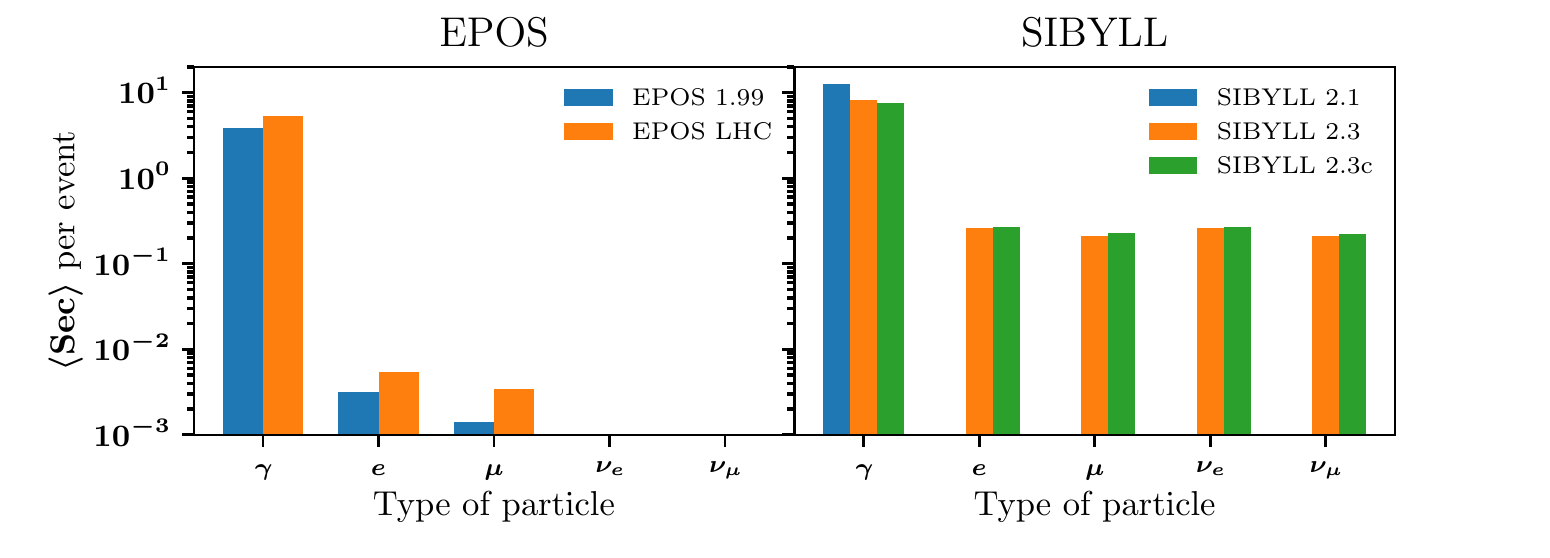}
	\caption{Distribution of secondary photons, leptons and neutrinos for the case of 100 EeV proton-nitrogen collisions.}
	\label{fig:SecDistpN100EeV_lepgam}
\end{figure}

Figures \ref{fig:SecDistpp56PeV_lepgam} and \ref{fig:SecDistpN100EeV_lepgam} display the average number of photons, leptons, and neutrinos produced during the EPOS or SIBYLL simulations. Notice that the number of such secondary particles produced are similar for all the versions of those models, except for the case of Fig. \ref{fig:SecDistpp56PeV_lepgam} where EPOS-LHC returns slightly more electrons and muons than EPOS 1.99. When comparing between models, it can be seen that both SIBYLL versions return approximately the same quantity of photons than EPOS, but a substantially larger number of leptons. Notice also that EPOS does not return neutrinos of any kind.

We continue our analysis by considering the differences between models in the deflection angles of the emerging secondary particles. In our previous work \cite{Gar09} we presented an exhaustive study of the secondary particle pseudorapidity $\eta$ ($\eta=-\ln[\tan(\theta/2)]$, where $\theta$ is the deflection angle of the corresponding secondary particle) distribution in several relevant cases. We found that the general characteristics of those distributions are maintained for the recent versions of all the models we considered, and for this reason, and for the sake of brevity we are not including all the details of our current analysis. 

We consider worthwhile presenting a comparison among the different pseudorapidity distributions in the case of proton-proton collisions at LHC energies. At this energy, all the models have been tuned against available experimental results in the range $|\eta|<2.5$ \cite{Kvita17}.

\begin{figure}[pt]
	\includegraphics{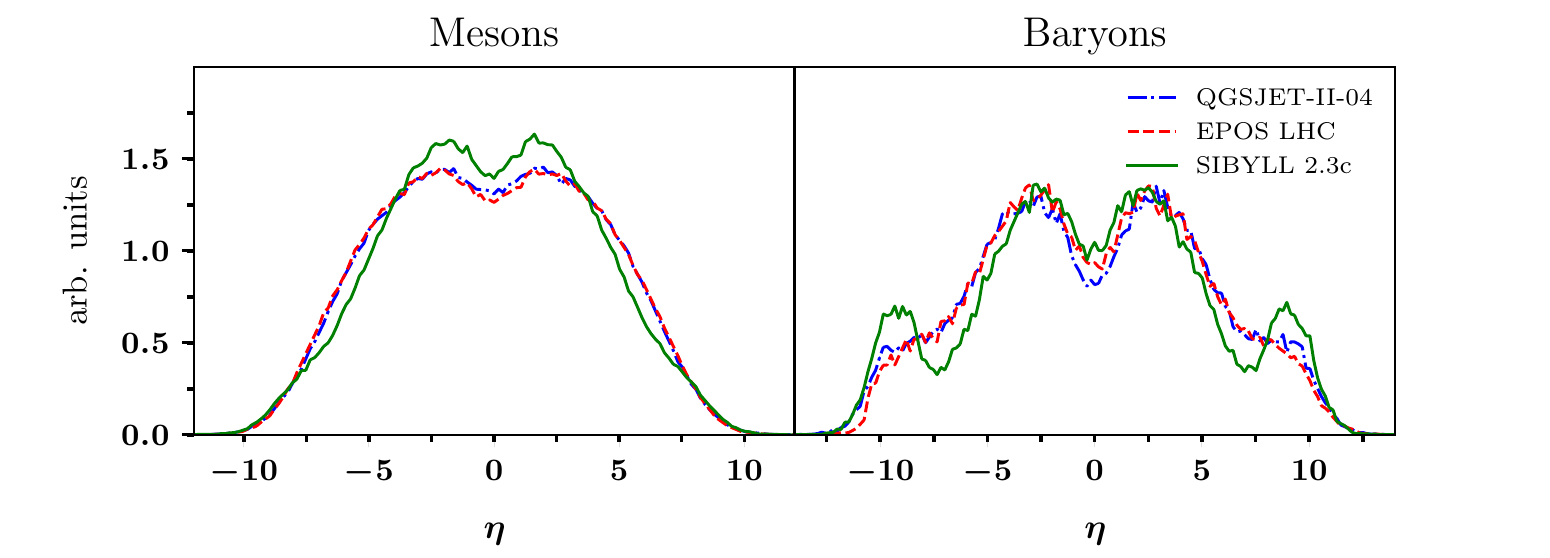}
	\caption{Normalized center of mass pseudorapidity distributions of the secondaries generated in 56 PeV proton-proton collisions.}
	\label{fig:CmEtaDist56PeVpp}
\end{figure}

In Fig. \ref{fig:CmEtaDist56PeVpp} we display the normalized center of mass pseudorapidity distribution in the case of 56 PeV (lab energy) proton-proton collisions. It shows up clearly that all the distributions present qualitatively similar shapes, especially in the region $|\eta|<2.5$. The similarity is particularly noticeable when comparing EPOS and QGSJET, and in this case extends to the whole range of $\eta$. On the other hand, the distributions for SIBYLL are somewhat different. In the case of mesons (mainly $\pi$, $K$, and $\rho$), they are slightly narrower than the corresponding ones for EPOS or QGSJET, thus implying that SIBYLL produces a moderately larger number of mesons in the pseudorapidity region that goes from -3 to 3 approximately. In the case of nucleons, there is a noticeable difference between the SIBYLL and the EPOS or QGSJET distributions, remarkable for $|\eta| \sim 10$ where the SIBYLL distribution presents visible peaks (see figure \ref{fig:CmEtaDist56PeVpp}).

\begin{figure}[pt]
	\includegraphics{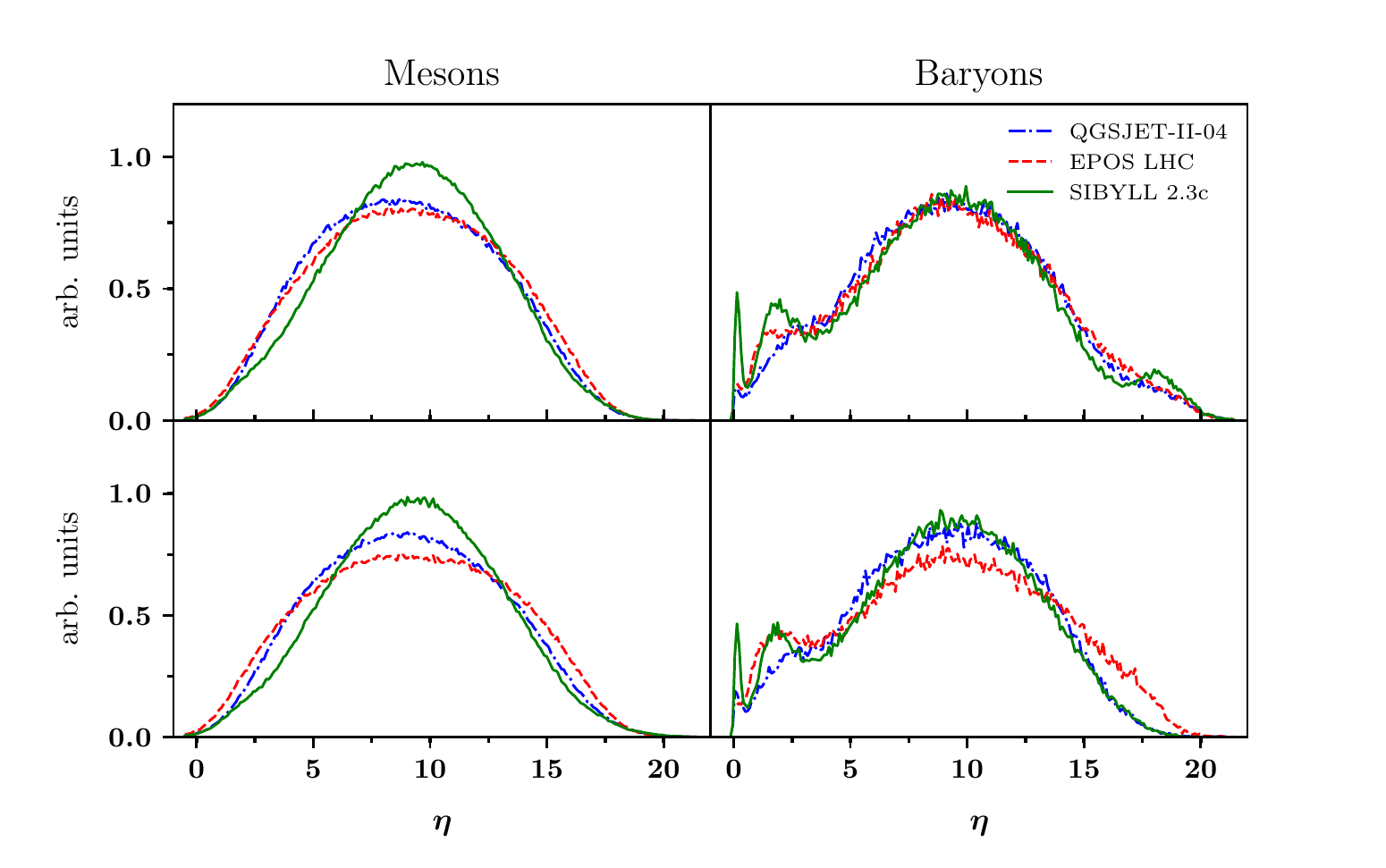}
	\caption{Normalized laboratory system pseudorapidity distributions of the secondaries generated in 56 PeV $p$-N (upper row) and $\pi^+$-N (lower row) collisions.}
	\label{fig:LabEtaDist56PeVpiNpN}
\end{figure}

In order to investigate possible effects on air shower development,
especially in the lateral distribution of particles, it is also
necessary to compare the pseudorapidity distributions in the lab
system, and in a typical case like collisions against nitrogen nuclei
targets. In Fig. \ref{fig:LabEtaDist56PeVpiNpN} we present such
distributions, always for the case of 56 PeV projectiles colliding
against nitrogen nuclei (for larger projectile energies all these
distributions are very similar in shape). In the case of mesons, the
distributions for all models are very similar, especially for EPOS and QGSJET. When considering the pseudorapidity of baryons, it
can be clearly seen that the different hadronic models lead to
noticeable differences, especially for $\eta$ between 0 and 4
($\theta$ between 2 and 90 degrees). The peak of the SIBYLL
distribution for proton-nitrogen collisions at $\eta \sim +18$ (upper
right plot in Fig. \ref{fig:LabEtaDist56PeVpiNpN}) corresponds to
VELP events characterized by a leading particle (nucleon) having a
very high energy and thus emerging with a very small deflection angle,
which corresponds to a large $\eta$. When the leading particle in VELP
events is not a baryon, as for the $\pi$-N collisions (lower right
plot in figure \ref{fig:LabEtaDist56PeVpiNpN}) such a peak is
absent. It is also noticeable that there is a relative increment of
baryons with $\eta \gtrsim 13$ for EPOS $\pi$-N collisions in
comparison with the other hadronic models. This is in agreement with the fact that as in EPOS diquarks are allowed as string ends, when the projectile is a meson, it leads to an increase of the (anti)baryon production in the forward direction \cite{Pierog:2017}. It is also important to remark that the baryon distributions present a sharp end at $\eta=0$ with no recoiling particles ($\eta<0$) being generated. This characteristic of the nucleons pseudorapidity distribution, also present in the old versions of the hadronic models \cite{Gar09}, is somewhat unnatural, and is accompanied by a relative abundance of particles with $\eta$ positive and very small, particularly noticeable in the case of SIBYLL.

\section{Impact on Shower Observables}
\label{sec:ShowerResults}

In this section we focus on the implications of the post-LHC updates in the hadronic interaction
packages on EAS observables that can provide information on the characteristics of the primary particle that initiates the shower. To achieve this goal, we have performed numerous sets of simulations with an updated version of the AIRES air shower simulation program \cite{Sci01} linked to every one of the hadronic models considered in this work.

One of the most important observables to consider is the shower maximum depth, $X_{\rm max}$ that is known to be proportional to the logarithm of the mass number of the primary particle. This observable has been extensively studied and the results are well-known (see, for example, Ref.\cite{Aab14a}). In our analysis of $X_{\rm max}$ obtained for all the hadronic packages considered here, we have reproduced such published results (not shown explicitly here for brevity). Comparing simulations with pre- and post-LHC models, it has been found that only SIBYLL presents an important change in $X_{\rm max}$ \cite{engel17}.

The signals associated with muons can be large enough to draw reliable conclusions on the primary mass or the possible appearance of new physics signatures, and for this reason most of the cosmic rays observatories are designed to be as efficient as possible to detect such kind of particles. Because of their importance, another set of observables that we have analyzed is related to the secondary muons produced during shower development.

The production of muons has been extensively discussed in the literature. In particular, it was found that the muon production simulated using the pre- and post-LHC hadronic models is significantly smaller when compared to the experimental observations and the differences among models are large \cite{Aab15}. Our analysis, not shown here for brevity, is in full agreement with the published  results.

The muon production depth (MPD) distribution is the other important observable,
because it can give valuable information about the primary mass \cite{Aab14b,Mallamaci17}. The MPD describes
the longitudinal development of the muonic component of the EAS and can be characterized by (a) its shape and (b) the point along the shower axis where the
production of muons reaches its maximum $X_{\rm max}^{\mu}$.  

\begin{figure}[pt]
	\includegraphics{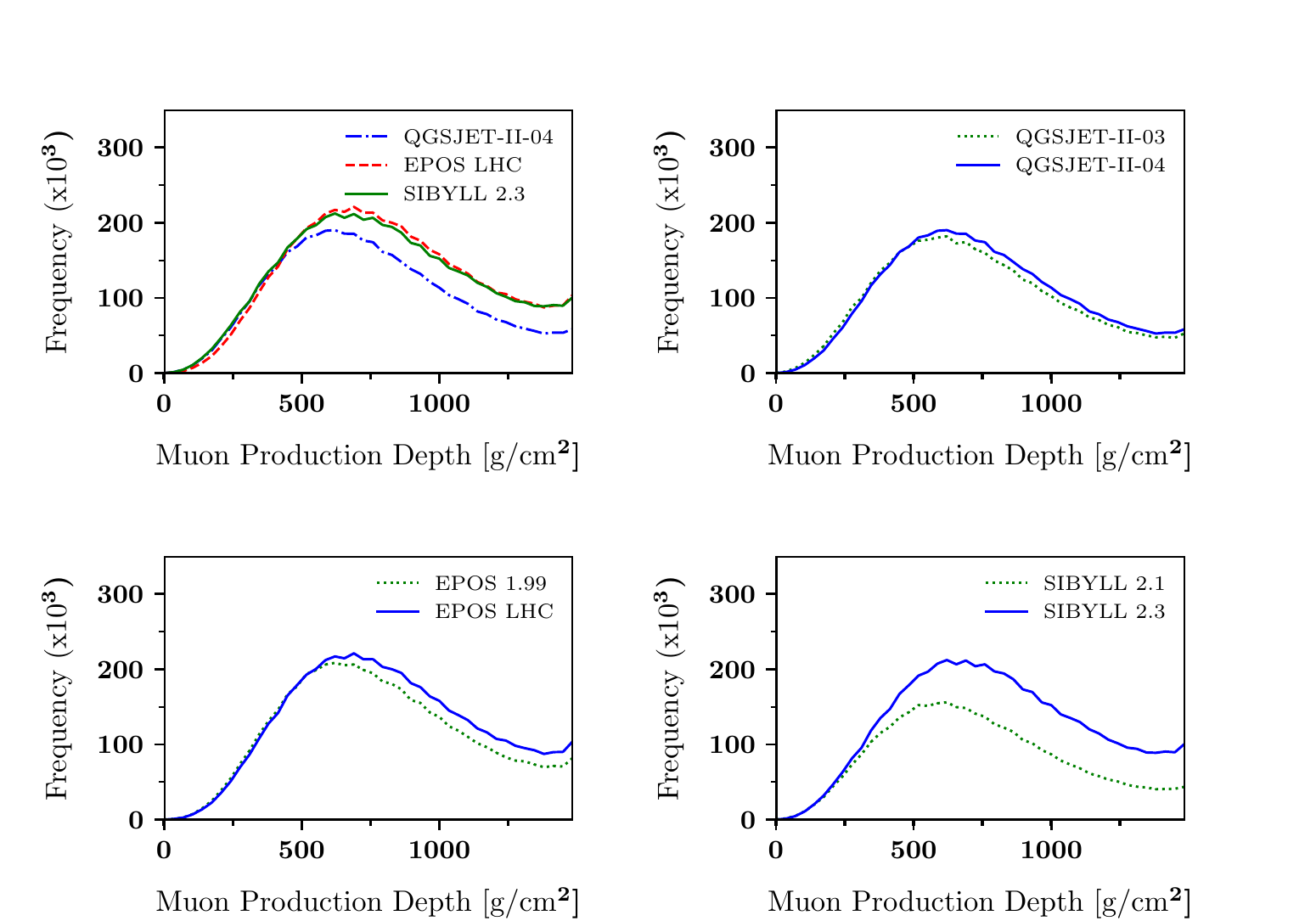}
	\caption{Muon production depth (MPD) distributions for showers
          simulated using the different hadronic models analyzed, in
          the case of 32 EeV proton showers inclined 55 degrees, and
          considering all muons that reach ground with kinetic energy
          greater than 60 MeV and distant more than 200 m from the
          shower axis.}
	\label{fig:MPDdist200mKemin060MeV}
\end{figure}

We have studied the MPD distribution in a series of representative cases, by means of air shower simulations. The results for the MPD distribution are plotted in Fig. \ref{fig:MPDdist200mKemin060MeV} for all the hadronic models considered, in the very representative case of 32 EeV proton initiated showers, inclined 55 degrees. 
From these plots it can be clearly seen that in the cases of EPOS and QGSJET there are no significant changes when comparing the corresponding pre- and post-LHC versions of these hadronic models. 
On the other hand, SIBYLL 2.3 (and 2.3c) produces, in comparison with the old version 2.1, a significantly larger number of muons and a slightly larger value of $X_{\rm max}^{\mu}$.
This is consistent with the fact that SIBYLL 2.3 gives noticeably larger values of $X_{\rm max}$, in comparison with SIBYLL 2.1, as discussed in Ref. \cite{Bell17}.

\begin{figure}[pt]
	\includegraphics{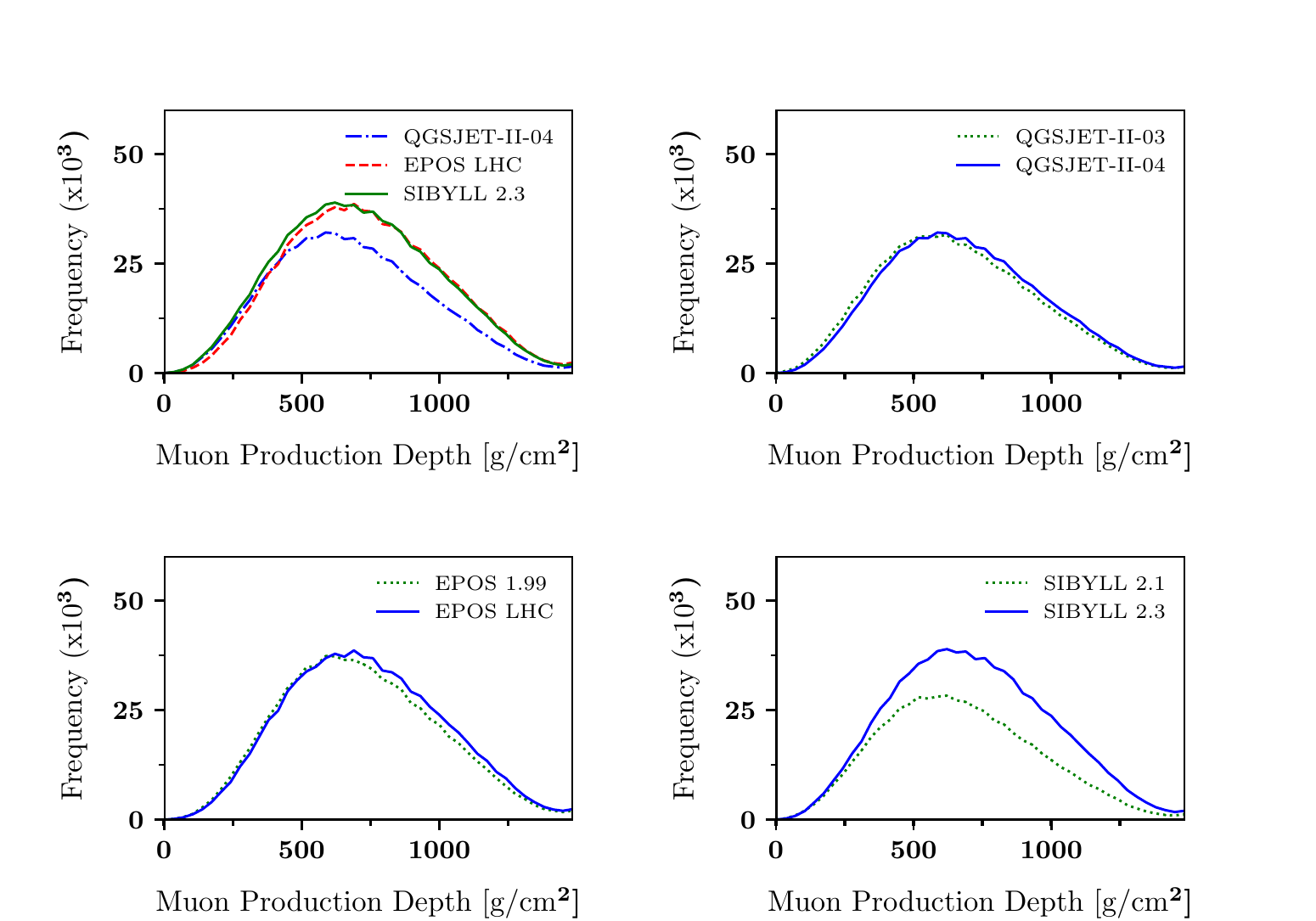}
	\caption{Same as figure \ref{fig:MPDdist200mKemin060MeV}, but
          considering all muons that reach ground at a distance from
          the shower axis that ranges between 1.2 and 4 km.}
	\label{fig:MPDdist1200mKemin060MeV}
\end{figure}

The MPD distribution depends on the subset of muons reaching ground that one considers for the analysis. This is clearly seen when comparing the plots in Fig. 
\ref{fig:MPDdist200mKemin060MeV} with the corresponding ones of Fig. \ref{fig:MPDdist1200mKemin060MeV}. Both figures represent the MPD, but for the cases of muons located more than $200$ m away from the shower axis (Fig. \ref{fig:MPDdist200mKemin060MeV}), or distant between $1200$ m and $4000$ m from such axis (Fig. \ref{fig:MPDdist1200mKemin060MeV}). The ``tails" of the distributions of Fig. \ref{fig:MPDdist200mKemin060MeV} correspond to muons produced very near the ground level after the decay of hadrons (mainly pions and kaons) that in turn are secondary particles of the hadronic collisions that took place near the shower axis. Such kind of decaying hadrons are much less frequent at larger distances from the shower axis, and thus the different shape of the distributions of Fig. \ref{fig:MPDdist1200mKemin060MeV}.

\section{Final remarks}
\label{sec:Final}

We have performed a comparative analysis of different observables associated with the secondary particles emerging from hadronic collisions simulated with pre- and post-LHC versions of the hadronic packages  SIBYLL, EPOS, and QGSJET.

Our analysis is focused on comparing pre- and post-LHC versions of each of the hadronic models studied in order to evaluate the changes between versions, and also in making comparisons among the newest versions of the models. We are particularly interested in describing the characteristics of the secondary particles produced by each model without entering in a detailed analysis of how such characteristics are related to the corresponding theoretical models. Such detailed analysis goes beyond the scope of the present work.

We have run an extensive set of simulations where a beam of hadronic projectile particles impacts on a given target generating a number of secondaries that were recorded and statistically analyzed. We considered proton and pion projectiles with energies ranging from 100 GeV to 300 EeV, as representative cases of the collisions that take place during the development of particle showers generated by cosmic rays that enter the Earth's atmosphere. The selection of nitrogen nuclei ($A=14$) as targets is also related with such atmospheric showers, since nitrogen is the most abundant component of air. Proton targets have also been used to allow for model comparisons in similar conditions as the collider experiments whose results were used to release improved versions of the simulation packages.

In many cases it is convenient to examine in detail the results of the simulations considering specific projectile and target, and with fixed projectile energy. In this work we have used very frequently two of such selections, namely, (1) proton-proton collisions with 56 PeV projectile kinetic energy in the laboratory system (corresponds to $\sqrt{s}\simeq 10$ TeV), as a representative case to analyze the results of the simulations in the conditions of the LHC experiments where data has been used to tune the newest versions of the hadronic models; and (2) proton-nitrogen collisions with 100 EeV projectile kinetic energy in the laboratory system, corresponding to the conditions of the highest energy cosmic ray collisions in the atmosphere, appropriate to study the results of the extrapolations used by the different models to simulate the collisions at energies well beyond the limits of present collider experiments.

The results presented in Sec. \ref{sec:Results} clearly indicate that there is a remarkable improvement in the degree of coincidence of many observables when comparing among similar simulations of the post-LHC versions of the three analyzed models. This is particularly evident in the case of the fractions of VELP events, that are approximately coincident for all the models studied and for all the primary energies considered in the case of proton-nitrogen collisions, as it clearly shows up in Fig. \ref{fig:FracVelppN} (upper left plot). Notice the remarkable difference between this plot and the respective one of our previous work (Fig. 4 of Ref. \cite{Gar09}).

Even if there is a good agreement among the fractions of VELP events returned by each of the studied models, the quantities that are closely related with that fraction, i.e., the inelasticity and the number of secondary particles, present moderate but non-negligible differences (see Figs. \ref{fig:Kdist100EeVpN}-\ref{fig:NsecpN100EeV} and the corresponding discussion in Sec. \ref{sec:Results}). Such differences occur even for the case of proton-proton collisions at energies comparable with the LHC ones.

Other characteristic of the hadronic collision simulators that has an important impact on the EAS development is the kind of generated hadrons, closely related to the production of other particles, particularly muons, after decays of the produced secondary hadrons. The analysis of the average number of different secondary hadrons produced by the hadronic models under identical conditions clearly show that there are significant differences among models. Such differences appear for most of the cases we have considered here, and include the case of proton-proton collisions at LHC energies. 

The analysis of EAS with different hadronic interaction packages reported in section \ref{sec:ShowerResults} indicates that the existing differences among hadronic models translates into differences in shower observables.

One of the most studied observables is the shower maximum depth, $X_{\rm max}$. The most recent comparisons with experimental data \cite{Aab14c,Aab17a,Bell17} indicate that predictions of simulations performed with different hadronic models using primary protons and atomic nuclei present a reasonable agreement with the measurements. It is possible to adequately reconcile experiment and simulations assuming that the cosmic rays hitting the Earth include protons and nuclei in proportions that vary with the primary energy. Under these conditions it is possible to determine the average mass, $\langle \ln A\rangle$, as a function of energy \cite{Bell17}, or, alternatively, to perform a simultaneous adjustment of the measured cosmic ray flux and its composition, assuming that the total flux is the sum of various components whose spectra are conveniently modeled and parameterized \cite{Aab17a}. In all cases, the simulations carried out with EPOS-LHC seem to be the ones that produce the best adjustments.

Another magnitude that has been measured by surface arrays such as the Auger Observatory is the number of ground-level muons produced by inclined showers . The results published in the reference \cite{Aab15}, as well as the simulations that we have done with all the versions (pre- and post LHC) of the hadronic models studied here, evidence without doubt that in all cases the number of muons predicted by the simulations is less than what results from the experimental measurements. In the representative case of 10 EeV showers, the relative deficit of simulated muons with respect to the measured ones is approximately 30\% (8\%) when compared with showers initiated by protons (iron nuclei) in the cases of SIBYLL 2.3 and EPOS-LHC. QGSJETII-04 predicts even fewer muons, 40\% and 23\% less, respectively. It is important to note that in the case of SIBYLL, there is a significant increase in the production of muons, which under the conditions of these simulations is of the order of 40\% larger compared to the old version 2.1. The pre- and post-LHC versions of EPOS and QGSJET do not show large variations with respect to the number of muons predicted in equal initial conditions. These results are consistent with the discussion presented in Secs. \ref{sec:Results} and \ref{sec:ShowerResults}.

Similarly, in the particularly interesting case of the MPD distribution we find that the LHC parameter tuning performed in the case of QGSJET and EPOS does not have a significant impact on this distribution; while in the case of SIBYLL there is a visible increase of the number of muons comparing the pre- and post-LHC versions, most probably related to a change in the fractions of secondary hadrons produced at each collision that allows for an increased muon production after the decay of unstable hadrons (mainly pions and kaons).

In all of the analyzed cases, the total number of produced muons continues to be significantly smaller in comparison with experimental measurements, regardless of the hadronic model used. The possibility of producing observables that are sensitive to the primary composition, such as, for example, quantities connected to the number of muons generated in the showers, as planned in the case of AugerPrime \cite{Aab16a}, will undoubtedly be of great importance as well as to improve our understanding of the nature of cosmic rays so as to be in a better position to validate the different hadronic models.

We end remarking that our analysis using different hadronic models allows us to conclude that there have been very significant improvements in the simulation of hadronic collisions, but this issue continues to be a challenging topic calling for further research, more than 30 years after the first simulations were reported, and despite all the theoretical efforts and the experimental data that have been collected since then.

\section*{ACKNOWLEDGMENTS}
This work was partially supported by Agencia Nacional de Promoción Científica y Tecnológica (ANPCyT), Argentina and Consejo Nacional de Investigaciones Científicas y Técnicas (CONICET), Argentina.
We are indebted to R. Clay, T. Pierog, and M. Unger for a careful reading of the manuscript and making useful comments.



\clearpage
\bibliography{aapaper1allv}

\clearpage


\end{document}